\documentclass[prodmode,acmtecs]{acmsmall} 

\usepackage[T1]{fontenc}
\usepackage{listings}
\usepackage[rgb,dvipsnames]{xcolor}
\usepackage{graphics}
\usepackage{array} 
\usepackage{afterpage} 
\usepackage{float} 
\usepackage{paralist} 
\usepackage[shortcuts]{extdash} 
\usepackage{todonotes}
\usepackage{textcomp}
\usepackage{ marvosym }
\usepackage{dirtree}
\usepackage{multirow,tabularx}
\usepackage{ctable}
\usepackage{relsize}
\usepackage{amsmath,amssymb}
\usepackage{pifont}
\usepackage{siunitx}
\usepackage[scaled=0.85]{beramono}
\usepackage[final]{microtype}
\microtypesetup{stretch=9,shrink=15,step=3,letterspace=50}

\usepackage[pass,letterpaper]{geometry}

\usepackage{balance}
\usepackage{subfig}
\usepackage{wrapfig}

\usepackage[ruled]{algorithm2e}

\SetAlFnt{\small}
\SetAlCapFnt{\small}
\SetAlCapNameFnt{\small}
\SetAlCapHSkip{0pt}
\IncMargin{-\parindent}

\begin{document}

\markboth{Matthias Springer}{Inter-language Collaboration in an Object-oriented Virtual Machine}

\title{Inter-language Collaboration in an Object-oriented Virtual Machine \\ \vspace{0.1cm} \small Bachelor's Thesis}
\author{MATTHIAS SPRINGER
\affil{Hasso Plattner Institute, University of Potsdam}
TIM FELGENTREFF
\affil{Hasso Plattner Institute, University of Potsdam (Supervisor)}
TOBIAS PAPE
\affil{Hasso Plattner Institute, University of Potsdam (Supervisor)}
ROBERT HIRSCHFELD
\affil{Hasso Plattner Institute, University of Potsdam (Supervisor)}
}

\begin{abstract}
Multi-language virtual machines have a number of advantages. They allow software developers to use software libraries that were written for different programming languages. Furthermore, language implementors do not have to bother with low-level VM functionality and their implementation can benefit from optimizations in existing virtual machines. MagLev is an implementation of the Ruby programming language on top of the GemStone/S virtual machine for the Smalltalk programming language. In this work, we present how software components written in both languages can interact. We show how MagLev unifies the Smalltalk and the Ruby object model, taking into account Smalltalk meta classes, Ruby singleton classes, and Ruby modules. Besides, we show how we can call Ruby methods from Smalltalk and vice versa. We also present MagLev's concept of bridge methods for implementing Ruby method calling conventions. Finally, we compare our solution to other language implementations and virtual machines.
\end{abstract}

%
%




\maketitle

\definecolor{greyBg}{rgb}{0.95,0.95,0.95}
\lstdefinestyle{customSmalltalk}{
  belowcaptionskip=1\baselineskip,
  breaklines=true,
  frame=lrbt,
  xleftmargin=\parindent,
  language=smalltalk,
  showstringspaces=false,
  basicstyle=\footnotesize\ttfamily,
  keywordstyle=\bfseries\color{green!40!black},
  commentstyle=\itshape\color{purple!40!black},
  stringstyle=\color{Maroon},
  moredelim=*[s][\ttfamily\color{gray}]{|}{|},
}
\lstset{  breaklines,
  breakatwhitespace}
\newcommand{\paragraphIndent}[1]{\paragraph{#1}\ }

\newcommand{\methRuby}{} 
\newcommand{\methSt}{} 

\section{Multi-language Virtual Machines}
Multi-language virtual machines are virtual machines that support the execution of source code written in multiple languages. According to Vran{\'y}~\cite{vranyj2010}, we have to distinguish multi-lanuguage virtual machines from interpreters and native code execution. Languages inside multi-language virtual machines are usually compiled to byte code and share the same object space. In contrast, an interpreter is a program that is written entirely in the host programming language and executes a guest programming language. The interpreter manages the communication between host and guest language, e.g. by converting objects and data types.

Multi-language virtual machines allow programmers to write parts of the program in different languages: they can ``use the right tool for each task''~\cite{Brunklaus02avirtual}. In the following paragraphs, we present the main advantages of multi-language virtual machines.

\paragraphIndent{Libraries}
Software libraries are a popular form of code reuse. They increase productivity~\cite{Basili:1996:RIP:236156.236184} and reduce software defects~\cite{Mohagheghi:2004:ESS:998675.999433}. By supporting multiple programming languages, we can use libraries that were written in different languages at the same time. Therefore, we do not have to reimplement functionality if we can find a library in one of the supported languages. The more programming languages are supported, the higher is the probability that we can find a library for our purposes.

JRuby\footnote{\url{http://jruby.org/}} is an implementation of the Ruby programming language on top of the Java Virtual Machine (JVM). Ruby libraries are usually packaged as so-called Gems. RubyGems\footnote{\url{http://rubygems.org/}} hosts more than 50,000 Gems and many of them can be used in JRuby. In addition, JRuby programmers can use Java libraries. Java is a popular and widely used programming language, and libraries are available for almost everything. MvnRepository\footnote{\url{http://mvnrepository.com/}} hosts more than 480,000 Java artifacts, however, also counting different version numbers.

\paragraphIndent{Performance and Portability}
A programming language that is implemented on top of an existing virtual machine can benefit from the virtual machine's performance and portability~\cite{conf/oopsla/MarrWHD11}. For example, programming languages for the Java Virtual Machine run on all operating systems and computer architectures that have a JVM implementation.

The JVM does a lot of optimizations and has a just-in-time compiler that compiles byte code to native code. Many programmers choose JRuby over the Ruby reference implementation (MRI) because, in most cases, it is faster~\cite{Cangiano:Shootout2008,Cangiano:Shootout2010}. Besides, JRuby supports parallel threads, i.e. it does not have a global interpreter lock~(GIL). 

\paragraphIndent{Language Implementation}
It is probably less work to implement programming languages on top of one performant virtual machine than developing a performant virtual machine from scratch for every programming language. Writing virtual machines is tedious because it involves a lot of low-level functionality like memory management, garbage collection, and just-in-time compilation. However, Vran{\'y} comments that ``current virtual machines were designed specifically for one programming language'' \cite{vranyj2010}. Furthermore, he states that virtual machines should be more open and language-independent to allow or simplify the implementation of new programming languages.

\paragraphIndent{Features provided by the Virtual Machine}
A programming language can expose all features that are provided by the virtual machine. For example, GemStone/S has a built-in object database. This database can be used from Smalltalk and MagLev, a Ruby implementation. The object database is seamlessly integrated into the programming language and is probably the main reason why programmers decide to use MagLev instead of other Ruby implementations.

\paragraphIndent{Outline of this Work}
The following sections describe how the Ruby programming language was implemented on top of GemStone/S. We put particular focus on language interaction concepts.
Section~\ref{sec:intro_maglev} gives a high-level overview of MagLev's architecture. The next sections describe three problems that occured during the implementation of the Ruby programming language in GemStone/S, and their solution in MagLev.
Section~\ref{sec:mapping_object_models} describes how the Ruby object model is mapped onto the Smalltalk object model, including singleton classes and modules.  
Section~\ref{sec:inter_lang_methods} describes how Ruby methods are implemented and how Ruby and Smalltalk methods can be called.  
Section~\ref{sec:instance_variables} explains how instance variables are accessed and implemented.
Section~\ref{sec:related-work} compares our solution of the presented problems and the implementation of MagLev to other implementations.
 
\section{Introduction to MagLev}
\label{sec:intro_maglev}
MagLev is an implementation of the Ruby programming language on top of GemStone/S. In this section, we explain important terms and definitions that we use throughout this work and give a short overview of MagLev.

\subsection{Programming Languages}

\paragraphIndent{Smalltalk} The Smalltalk programming language is an object-oriented, dynamically typed programming language that was designed by Alan Kay, Dan Ingalls, and Adele Goldberg at Xerox PARC. It was a popular programming language in the mid-1990s. Smalltalk-80~\cite{Goldberg1983} is a specification of the Smalltalk programming language.

\paragraphIndent{Ruby} The Ruby programming language is an object-oriented, dynamically typed programming language that was developed by Yukihiro Matsumoto. It became popular with the web application framework Ruby on Rails. Ruby MRI (\emph{Matz's Ruby Interpreter}) is the reference implementation. All other Ruby implementations, such as Rubinius, JRuby, and MagLev, try to be as compatible as possible to Ruby MRI. Ruby has many similarities with Smalltalk and some people even see Ruby as Smalltalk's successor. The inventor of Extreme Programming, Kent Beck, once said ``I always knew that one day Smalltalk would replace Java. I just didn't know it would be called Ruby''~\cite{Bowkett:Smalltalk:2007}.

\subsection{Terms and Definitions}

\paragraphIndent{GemStone/S} GemStone/S is a Smalltalk specification that is very close to the Smalltalk-80 standard. In contrast to Smalltalk implementations like Pharo and Squeak, it does not come with a graphical user interface (e.g. Morphic). GemStone/S has a builtin object database that persists objects \emph{living} in the image~\cite{Butterworth:1991:GOD:125223.125254,Maier:1986:DOD:28697.28746}. Smalltalk source code is compiled to byte code that is then executed by the GemStone/S virtual machine.

\paragraphIndent{MagLev} MagLev is an implementation of the Ruby programming language on top of GemStone/S. Its object persistence concepts are integrated in Ruby seamlessly. MagLev currently supports Ruby~1.8 and there is experimental support for Ruby~1.9.

\paragraphIndent{Environments} For the implementation of MagLev, GemStone/S was changed in such a way that it can support multiple programming languages through \emph{environments}. We can think of environments as enclosed parts of the system where only one programming language is allowed. GemStone/S has a Ruby and a Smalltalk environment.

\paragraphIndent{Class Names}
In MagLev, classes can have different names in the Ruby environment and the Smalltalk environment. Whenever we are referring to a class with its Ruby name, the name is prefixed with two colons. For example, \lstinline{::Hash} is a Ruby class name but \lstinline{RubyHash} is a Smalltalk class name.

\subsection{Basic Concepts of MagLev}
\begin{figure}
\begin{center}\includegraphics[]{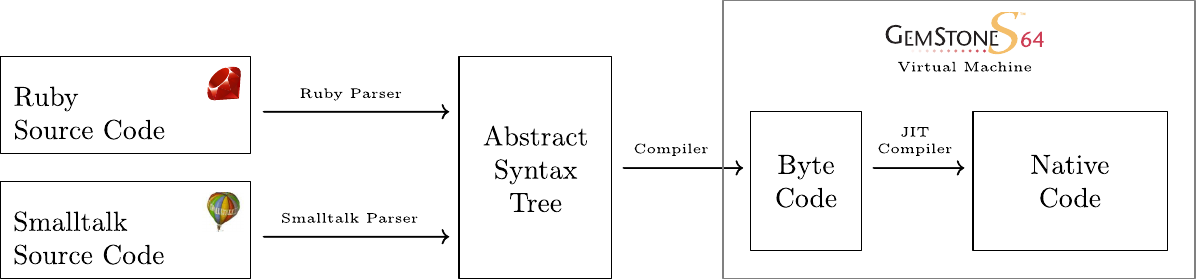}\end{center}
\caption{Integration of MagLev in GemStone/S' architecture.}
\label{fig:maglev_architecture}
\end{figure}

Figure~\ref{fig:maglev_architecture} shows a high-level overview of MagLev's architecture. Ruby source code and Smalltalk source code is transformed into an abstract syntax tree (AST). The compiler converts the AST to byte code. GemStone/S' just-in-time compiler eventually generates native code and executes it.

In MagLev, every Ruby object is a Smalltalk object and vice versa. This is also true for classes. It does not matter in what programming language a method was written. Both Ruby and Smalltalk code is compiled to byte code and can call methods that were written in the other language. We can also see MagLev as a compatibility layer for Ruby built on top of GemStone/S. It adjusts the Smalltalk object model to comply with the Ruby object model. Furthermore, MagLev provides the complete Ruby 1.8 standard library that is written in Ruby but reuses existing functionality provided by GemStone/S. 

\section{Mapping Object Models}
\label{sec:mapping_object_models}
MagLev does not distinguish between Ruby objects and Smalltalk objects: they are the same. Both languages are purely object-oriented, i.e. everything is an object. However, Smalltalk's meta class model differs from Ruby's singleton class model. Furthermore, Ruby supports mixins, whereas GemStone/S does not. In this section, we describe how MagLev maps Ruby's and Smalltalk's object model, such that we can use the same objects in the Ruby environment and in the Smalltalk environment. Most importantly, existing applications should still work correctly, even if they make assumptions about the underlying object model.

\subsection{Classes in Ruby and Smalltalk}
In Smalltalk, classes have an association entry in the globals dictionary. We can mention the class names in the Smalltalk source code and the compiler automatically replaces them by references to the class. Ruby has a different concept: classes and modules define namespaces~\cite{Bergel05analyzingmodule}. They are added as constants to another class or module. The root of the namespace hierarchy is \lstinline{::Object}. MagLev makes Smalltalk classes available to the Ruby environments by adding constants with a reference to the class object to \lstinline{::Object}. Afterwards, we can reference these classes in Ruby. 

MagLev can map most Smalltalk classes directly to Ruby classes, e.g. \lstinline{Object} $\leftrightarrow$ \lstinline{::Object}, \lstinline{String} $\leftrightarrow$ \lstinline{::String}, \lstinline{UndefinedObject} $\leftrightarrow$ \lstinline{::NilClass}. Most classes in Ruby's standard library are implemented completely in Ruby and are not known in the Smalltalk environment, e.g. \lstinline{::WEBrick::HTTPServer}, an HTTP web server. There are also some Smalltalk classes that are not known in the Ruby environment, e.g. AST classes and GCI\footnote{The GemBuilder for C Interface (GCI) is used to communicate with a running GemStone/S image via network.} classes.

It is important to remember that all Ruby objects are Smalltalk objects and vice versa. Therefore, this is also the case for all classes. Some classes simply do not have a Smalltalk name or a Ruby name, making it harder to get a reference. If we know a class' object id, we can always retrieve the class object and do whatever we want to, both in Ruby and in Smalltalk.

\subsection{Ruby Singleton Classes and Smalltalk Meta Classes}
Ruby and Smalltalk have related concepts of singleton classes and meta classes, respectively some people call Ruby's concept of singleton classes a more consequent and object-oriented implementation of Smalltalk's meta classes~\cite{pavlata:rubyObjectModel}.

\paragraphIndent{Smalltalk} Every Smalltalk class is an instance of its own meta class~\cite{Goldberg1983} that contains the methods and instance variables for the class side. The meta class is automatically generated for every non-meta class, and its superclass hierarchy is parallel to its non-meta class' superclass hierarchy.


\begin{figure}
\begin{center}\includegraphics[scale=0.54]{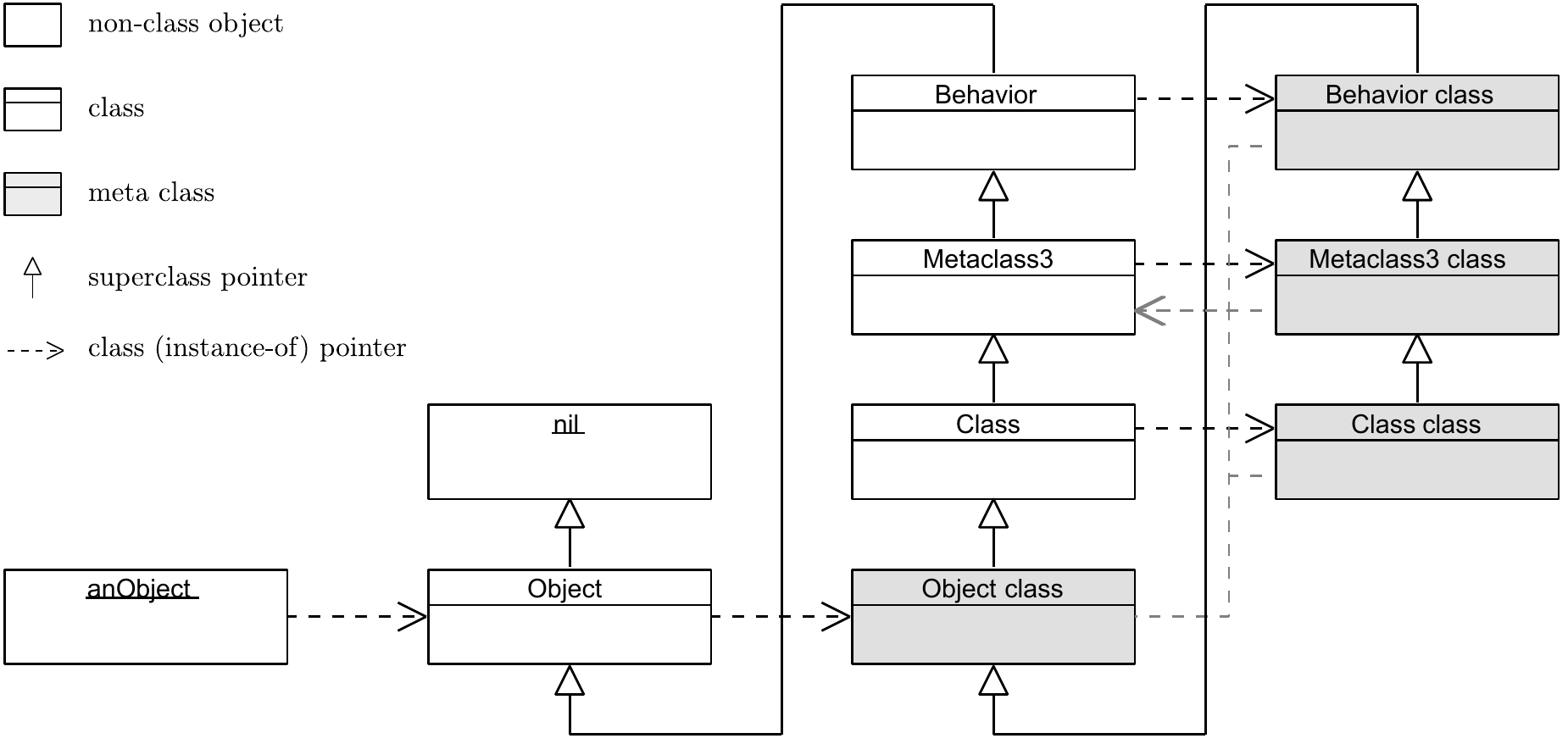}\end{center}
\caption{GemStone/S' meta class model. Meta classes are colored gray and instances of \lstinline{Metaclass3}.}
\label{fig:prob_classmodel_smalltalk_meta}
\end{figure}

Figure~\ref{fig:prob_classmodel_smalltalk_meta} shows a part of GemStone/S' object model. All meta classes are instances of \lstinline{Metaclass3}. Notably, \lstinline{Metaclass3 class} is also an instance of \lstinline{Metaclass3}.

\begin{figure}
\begin{center}\includegraphics[scale=0.54]{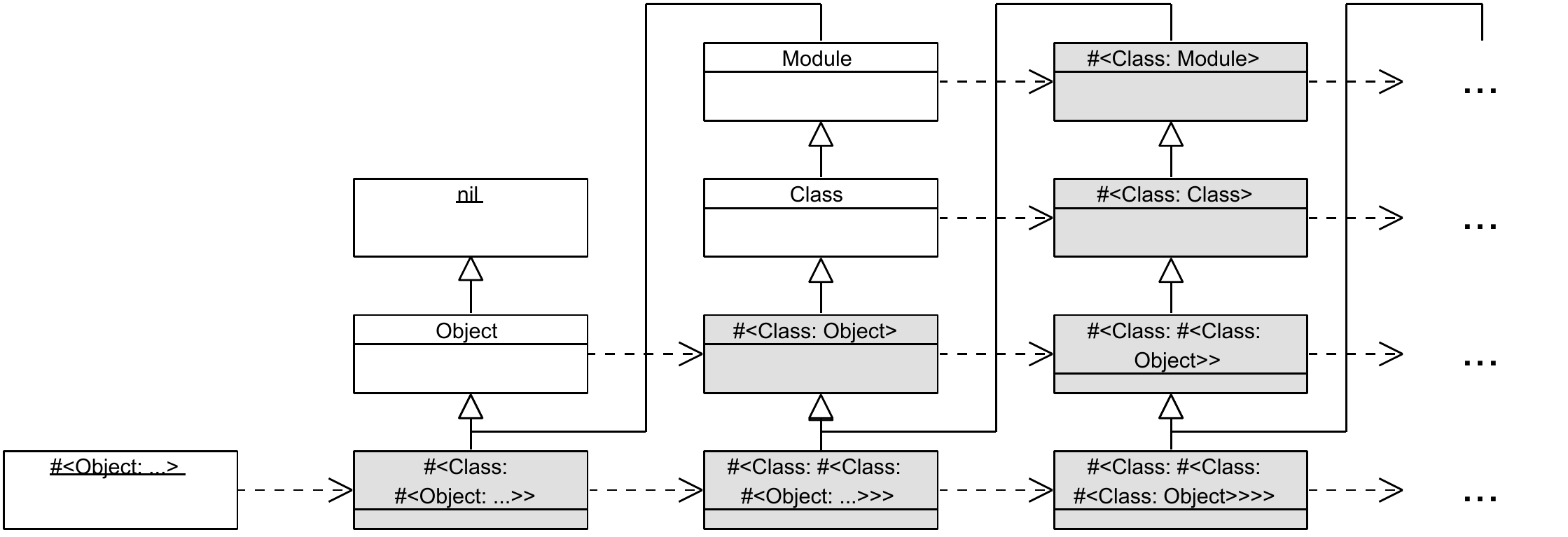}\end{center}
\caption{Ruby's singleton class model in version 1.8. In Ruby 1.9, \lstinline{Object} is a subclass of \lstinline{BasicObject}, which is a subclass of \lstinline{nil}. Singleton classes are colored gray. \lstinline{\#<Class: anObject>} is the Ruby notation for \lstinline{anObject class}.}
\label{fig:prob_classmodel_ruby_singleton}
\end{figure}

\paragraphIndent{Ruby} The Ruby programming language has a more complex object model. Instead of meta classes, Ruby has the concept of singleton classes (also called eigenclasses). Every object is an instance of its singleton class (Figure~\ref{fig:prob_classmodel_ruby_singleton}). Therefore, every object can have its own methods that are not available for other objects of the same class. Just as meta classes, singleton classes can have only one instance. In Smalltalk terms, a non-singleton class' first level singleton class can be seen as its meta class. However, singleton classes are instances of their own singleton classes, as well, whereas meta classes are always instances of \lstinline{Metaclass3} in Smalltalk.

The classes \lstinline{Object}, \lstinline{Module}, and \lstinline{Class} are called helix classes~\cite{pavlata:rubyObjectModel} because they are instances of themselves. For example, \lstinline{Class}' class is \lstinline{#<Class: Class>} which is an indirect subclass of \lstinline{Class}. With Ruby 1.9, a fourth helix class, \lstinline{BasicObject}, was introduced.

\subsection{Problem} In MagLev, every Smalltalk object is a Ruby object and vice versa. Many popular Rubygems, such as Ruby on Rails Gems, use singleton classes intensively, making it an important language feature~\cite{gunter:metaprog}. The following features must be supported by MagLev and the GemStone/S virtual machine.
 
\begin{itemize}
  \item Generating singleton classes. GemStone/S supports only first-level singleton classes (meta classes). MagLev must be able to generate higher-level singleton classes.
  \item Singleton class-aware method lookup. For example, an instance method defined in \lstinline{#<Class: Object>} must be callable from instances of \lstinline{#<Class: Class>}.
  \item Interacting with singleton classes in Smalltalk. We need a way to access singleton classes in Smalltalk and call methods on them.
  \item Compatibility to exisiting Smalltalk code. The class hierarchy should not be changed heavily because some Smalltalk applications might make assumptions about instance-of and superclass relations.
\end{itemize}

\subsection{Solution}
In Ruby, every object is an instance of its singleton class. Smalltalk supports first-level singleton classes (meta classes) only. Our approach combines the concept of singleton classes with Smalltalk's object model.

\begin{figure}
\begin{center}\includegraphics[scale=0.54]{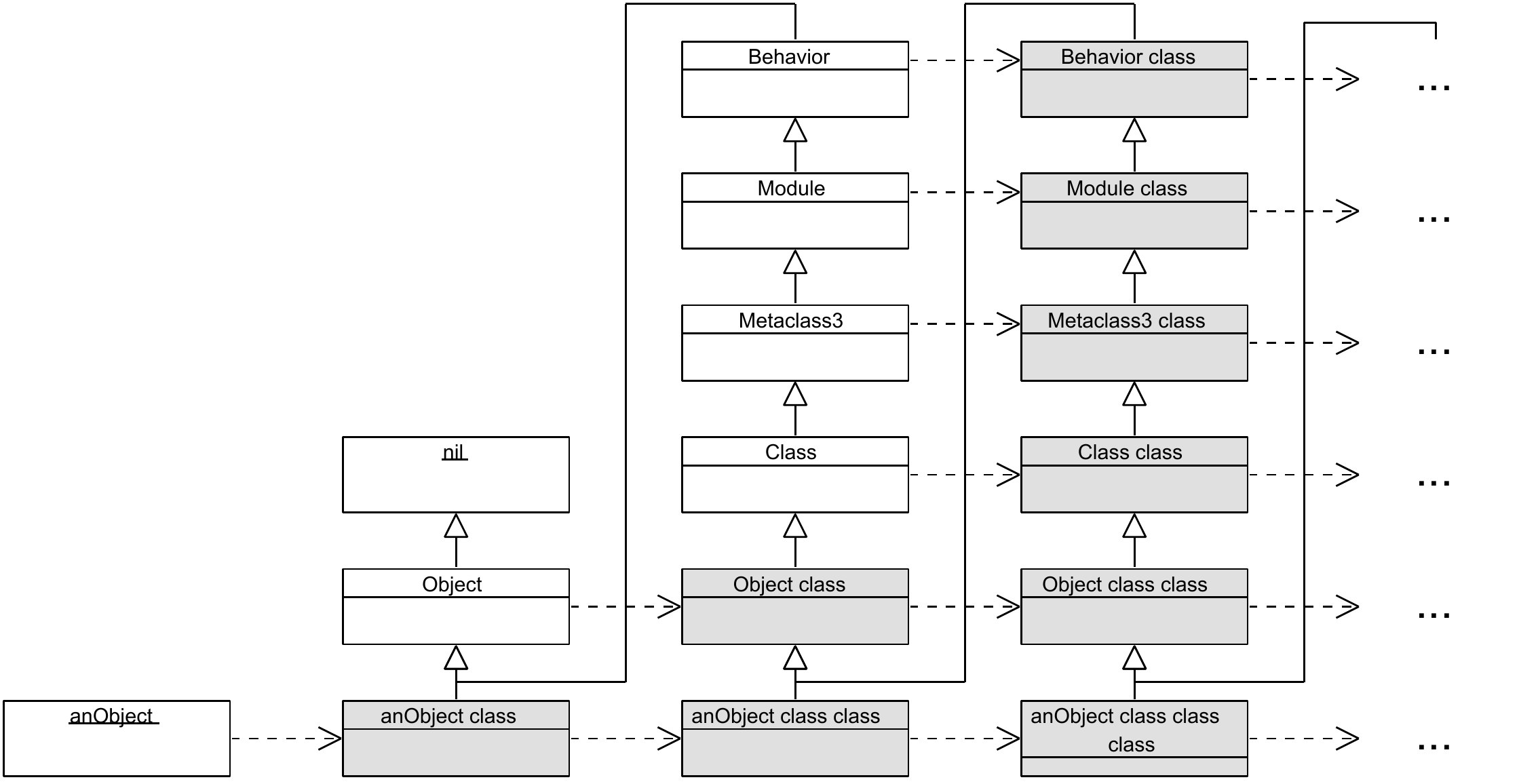}\end{center}
\caption{Combining Smalltalk's and Ruby's object model with singleton classes, resulting in five helix classes. Singleton classes are colored gray. The class \lstinline{Module} is needed for Ruby modules only and not relevant at this point.}
\label{fig:prob_classmodel_maglev_singleton}
\end{figure}

Figure~\ref{fig:prob_classmodel_maglev_singleton} shows what the new object model looks like from the programmer's point of view. Every object is an instance of its own singleton class. This horizontal instance-of relationship repeats forever.

\paragraphIndent{Generating Singleton Classes} The virtual machine cannot generate all singleton classes on class creation because an infinite number of singleton class levels exists for every class. Therefore, we only generate higher-level singleton classes when they are accessed for the first time. The first-level singleton class (Smalltalk meta class) is, however, automatically generated for Smalltalk compatibility reasons.

\begin{figure}
  \begin{equation}
    \mathit{superclass}(\mathit{singleton}(\mathit{obj})) = 
    \begin{cases}
      \mathit{class}(\mathit{obj}) & \text{if $\mathit{obj}$ is not a class} \\
      \texttt{Class} & \text{if $\mathit{obj} == \texttt{Object}$} \\
      \mathit{singleton}(\mathit{superclass}(\mathit{obj})) & \text{else}
    \end{cases}
  \nonumber
  \end{equation}
  \caption{Computing the superclass for $obj$'s singleton class.}
  \label{fig:eq:superclass_singleton_obj}
\end{figure}


When a singleton class for the object $obj$ is accessed for the first time, we have to generate it by subclassing from another class. In Figure~\ref{fig:prob_classmodel_maglev_singleton} we can see that a singleton class' superclass is, in most cases, the superclass' singleton class. Figure~\ref{fig:eq:superclass_singleton_obj} shows the formula for computing a singleton class' superclass. It contains special cases for \lstinline{Object} and non-class objects because these objects do not have a superclass. If the class that was computed by the formula does not exist, we have to generate it first, using this algorithm recursively.

After we generated the singleton class for an object, we set the object's class pointer to its singleton class. Singleton classes are always instances of \lstinline{Metaclass3} until we generate the next-level singleton class.

\paragraphIndent{Generating two Levels of Singleton Classes} Whenever we invoke a method on an object, the virtual machine searches for the selector in the method dictionary of the object's class. Therefore, we must additionally generate the second-level singleton class when the first-level singleton class is accessed for the first time, in order to assure that class methods work correctly on singleton classes.

Consider, for instance, that the method \methRuby\lstinline{example} is defined on \lstinline{Object class}. We should be able to call this method by sending \lstinline{example} to \lstinline{Object class class}. Therefore, we need to make sure that \lstinline{Object class class class} exists and is a subclass of \lstinline{Object class}.

\subsection{Example}
\begin{figure}
\begin{center}\includegraphics[scale=0.54]{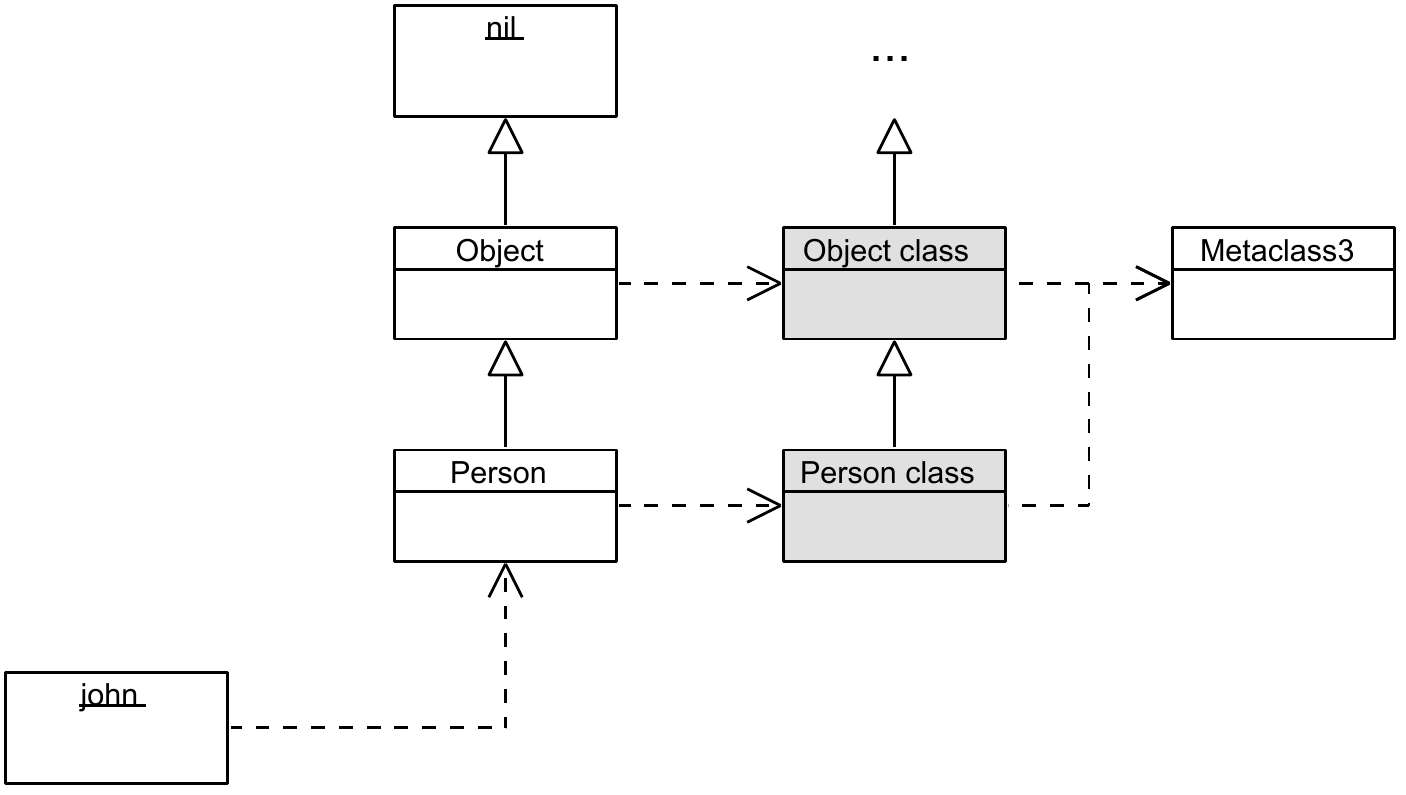}\end{center}
\caption{Creating the class \lstinline{Person} and its instance \lstinline{john}.}
\label{fig:singleton_class_create1}
\end{figure}

In this example, we show how MagLev generates the second-level singleton class for the object \lstinline{john} that is instance of the class \lstinline{Person}, without generating two levels of singleton classes. Figure~\ref{fig:singleton_class_create1} shows the situation after we created the class \lstinline{Person} and the instance \lstinline{john}. GemStone/S automatically generated the first-level singleton class, that is an instance of \lstinline{Metaclass3}.

\begin{figure}
  \centering
      \subfloat[Generating \lstinline{john}'s first-level singleton class. The first case in Figure~\ref{fig:eq:superclass_singleton_obj} applies.]{\includegraphics[scale=0.54]{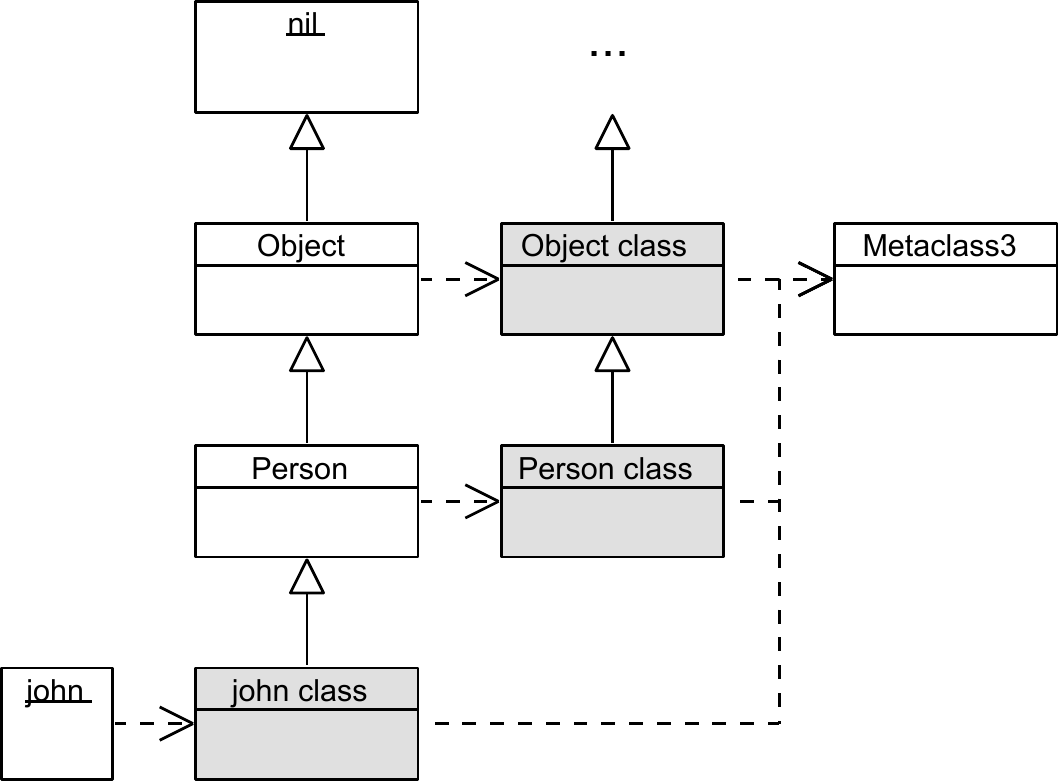}\label{fig:singleton_class_create2}} \hfill
      \subfloat[Generating \lstinline{john}'s second-level singleton class. The third case in Figure~\ref{fig:eq:superclass_singleton_obj} applies.]{\includegraphics[scale=0.54]{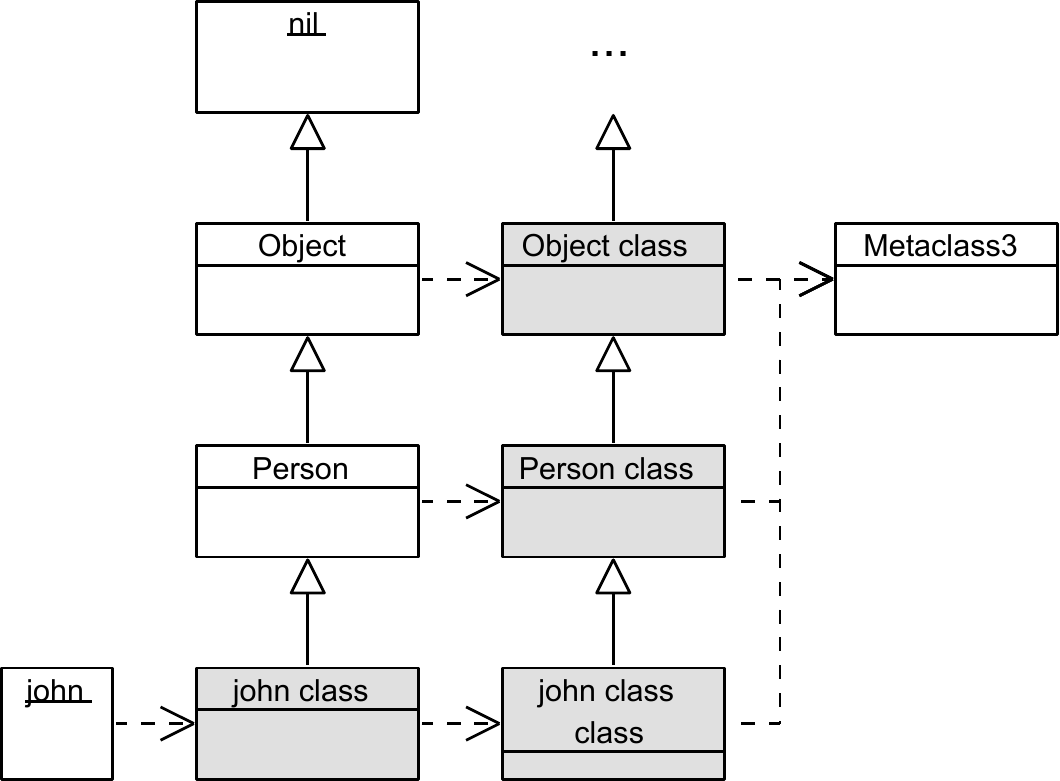}\label{fig:singleton_class_create3}}
   \caption[First-level and second-level singleton classes]{Generating \lstinline{john}'s first-level and second-level singleton class. We do not need to generate more singleton classes recursively.}
   \label{fig:singleton_class_create23}
\end{figure}

Now, we want to generate \lstinline{john}'s first-level singleton class (Figure~\ref{fig:singleton_class_create2}). According to Figure~\ref{fig:eq:superclass_singleton_obj}, the singleton class' superclass is \lstinline{Person} because \lstinline{john} is not a class. Here, we see that it is important to generate two levels of singleton classes: we cannot call methods defined in \lstinline{Person class} on \lstinline{john class} although this should be possible. \lstinline{john class} is still an instance of \lstinline{Metaclass3} that does not have instance methods that were defined in \lstinline{Person class}.

For \lstinline{john}'s second-level singleton class (Figure~\ref{fig:singleton_class_create3}), the superclass is \lstinline{Person class}. In both cases, the superclass already exists, so no further work needs to be done.

\begin{figure}
\begin{center}\includegraphics[scale=0.54]{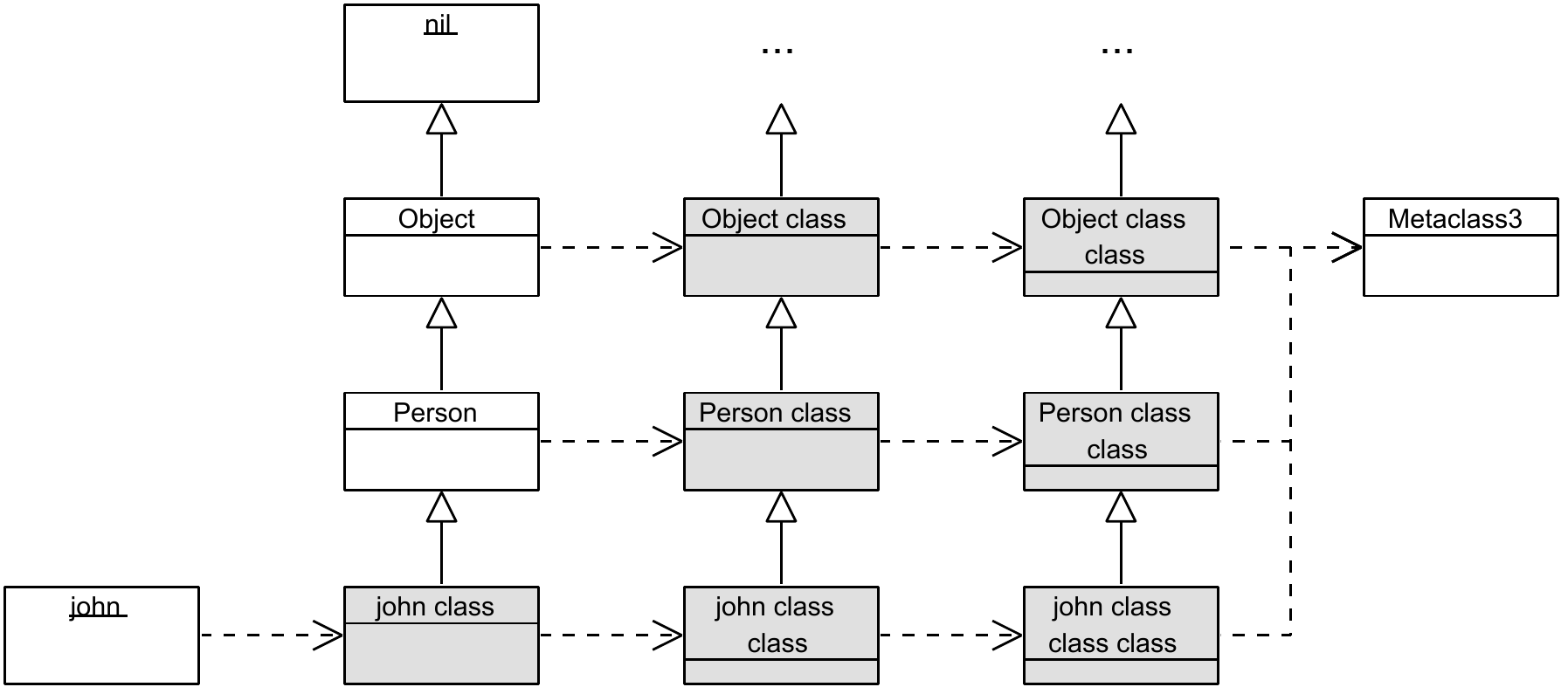}\end{center}
\caption{Generating \lstinline{john}'s third-level singleton class. More singleton class must be generated recursively.}
\label{fig:singleton_class_create4}
\end{figure}

If we want to generate \lstinline{john}'s third-level singleton class (Figure~\ref{fig:singleton_class_create4}), its superclass should be \lstinline{Person class class}. This class does not exist yet. Therefore, we generate \lstinline{Person class}' singleton class recursively. \lstinline{Person class class}' superclass does, again, not exist yet. In the end, we have to generate second-level singleton classes for \lstinline{Person} and \lstinline{Object}, in addition to \lstinline{john}'s third-level singleton class. First-level singleton classes for all helix classes already exist.

\subsection{Implementation}
In this section, we describe some chracteristics of the implementation of singleton classes in MagLev.

\paragraphIndent{Terms and Definitions}
The following list explains some terms that we use in the context of GemStone/S and MagLev throughout the implementation sections and the code samples.

\begin{figure}
\begin{center}\includegraphics[scale=0.54]{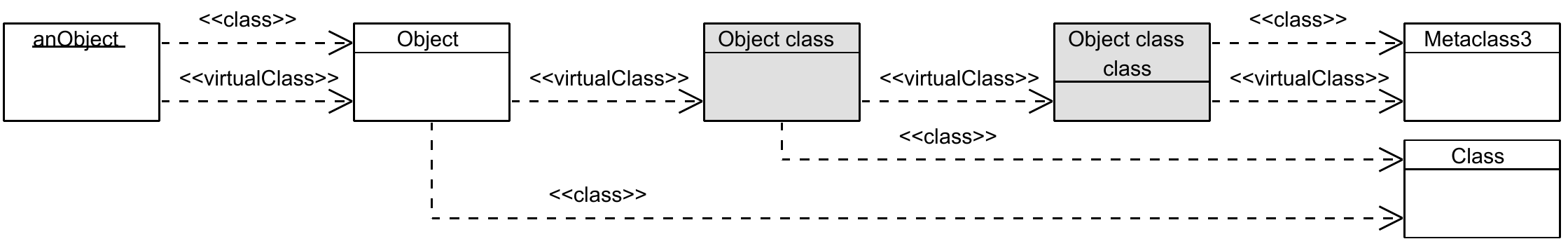}\end{center}
\caption{Example for terms and definitions. Meta classes are colored gray. \lstinline{anObject} does not have a singleton class generated, neither does \lstinline{Object class class}.}
\label{fig:class_virtual_class}
\end{figure}

\begin{itemize}
  \item An object's \emph{Virtual Class}: the class that is connected to the object by the instance-of relation.
  \item A class is a \emph{Meta Class} if it is a Smalltalk meta class or a Ruby singleton class. First-level singleton classes of non-meta classes are \emph{Smalltalk Meta Classes}. For example, \lstinline{Object class} is a meta class and a Smalltalk meta class. \lstinline{Object class class} is a meta class.
  \item An object's \emph{Class}: the first non-meta superclass in the virtual class' superclass hierarchy. For example, in Figure~\ref{fig:singleton_class_create4}, \lstinline{john}'s class is \lstinline{Person} and \lstinline{john class class}' class is \lstinline{Class}. An object's virtual class is equal to its class iff no singleton class was generated for the object, yet. See Figure~\ref{fig:class_virtual_class} for some examples.
  \item A meta class' \emph{Destination Class} is the object whose virtual class is the meta class. Therefore, it is the inverse virtual class relation.
\end{itemize}

\paragraphIndent{Objects and Classes in the GemStone/S Virtual Machine} In the GemStone/S virtual machine, every object is internally represented by a C++ object. Every object has an object ID (oop), flags, and a pointer to its virtual class object. The virtual class pointer is used during method lookup.

Classes are always instances of a class that inherits from \lstinline{Behavior} and \lstinline{Metaclass3}. \lstinline{Behavior} provides the instance variable \lstinline{format} that contains a 32-bit Integer with flags for the class. One of these flags determines if the class is a meta class and can be set in Smalltalk and in the virtual machine. \lstinline{Metaclass3} provides the instance variable \methSt\lstinline{destClass} that references the destination class. Among other things, the destination class pointer is necessary to generate a singleton class' name.

\paragraphIndent{Generating Singleton Classes}
In MagLev, we generate singleton classes when they are accessed for the first time. This happens when the programmer calls the method \methRuby\lstinline{singleton_class} or when the Ruby compiler operates on the singleton class while traversing the abstract syntax tree. In both cases, the method \methSt\lstinline{rubySingletonClass} is called to retrieve the singleton class object.

\methSt\lstinline{rubySingletonClass} calls \methSt\lstinline{ensureSingletonClassGenerated: 2} to make sure that two levels of singleton classes are generated. Then the virtual class is returned. Figure~\ref{lst:ensure_singleton_class_generated} shows how this method recursively generates multiple levels of singleton classes.

\begin{figure}
  \lstset{escapeinside={@}{@}}
  \begin{lstlisting}
@\textbf{Object}@>>ensureSingletonClassGenerated: depth
  depth == 0 ifTrue: [^ self].
  self checkGenerateSingletonClass.
  self virtualClass ensureSingletonClassGenerated: depth - 1.
  \end{lstlisting}

  \caption{Entry point for singleton class generation.}
  \label{lst:ensure_singleton_class_generated}
\end{figure}

We can distinguish two cases in which a singleton class must be generated. In both cases, the virtual class is not a meta class\footnote{Keep in mind that singleton classes are also considered meta classes.} (Figure~\ref{fig:check_generate_singleton_class}).

\begin{itemize}
  \item The object is not a class and no (first-class) singleton class was generated, yet. In that case, the virtual class is the class of the object, which is not a meta class.
  \item The object is a singleton class and no higher-level singleton class was generated, yet. In that case, the virtual class is \lstinline{Metaclass3}, which is not a meta class.
\end{itemize}

\begin{figure}
  \lstset{escapeinside={@}{@}}
  \begin{lstlisting}
@\textbf{Object}@>>checkGenerateSingletonClass
  self singletonAllowed 
    ifFalse: [^ self].
  self virtualClass isMeta
    ifFalse: [self generateSingletonClass].
  \end{lstlisting}

  \caption{Generating a singleton class, if it does not yet exist. In GemStone/S, we do not generate singleton classes for some special Smalltalk classes. This is an implementation detail and not discussed in this work.}
  \label{fig:check_generate_singleton_class}
\end{figure}

The method \methSt\lstinline{generateSingletonClass} is responsible for creating \lstinline{self}'s singleton class. This involves some operations that must be done in the lower-level VM code. The following list shows how the singleton class is generated.

\begin{enumerate}
  \item Compute the singleton class' superclass according to Figure~\ref{fig:eq:superclass_singleton_obj}. This might involve generating more singleton classes if the superclass does not exist, yet.
  \item Generate a new class that is instance of \lstinline{Metaclass3} and set the superclass.
  \item Set the meta class bit.
  \item Set the instance variable \methSt\lstinline{destClass} (destination class) to \lstinline{self}.
  \item Set \lstinline{self}'s virtual class pointer to the newly-generated singleton class.
\end{enumerate}


\subsection{Evaluation}

Our implementation solves the problems we identified previously: we can generate as many levels of singleton classes as we want to. Singleton classes are considered during method lookup because they are inserted into the class hierarchy as superclasses. From the Smalltalk side, we can access existing singleton classes with \lstinline{virtualClass} and generate and access singleton classes with \lstinline{rubySingletonClass}.

\paragraphIndent{Performance Issues} GemStone/S does some security checks in the virtual machine. For example, the implementation of \lstinline{isKindOf:} ensures that  the argument or the argument's class is an instance of \lstinline{Metaclass3}. With singleton classes, this is not necessarily the case anymore. However, the singleton class' superclass is always a subclass of \lstinline{Metaclass3}. Therefore, we need to check the whole superclass hierarchy. This is slower than the original implementation.

\paragraphIndent{Recursive Singleton Class Generation}
If a singleton class' superclass does not exist, it is generated recursively. This might require generating even more superclasses. In Figure~\ref{fig:eq:max_singleton_classes}, $\mathit{maxGen}(\mathit{obj})$ is the worst-case number of singleton classes that must be generated when accessing \emph{obj}'s singleton class.

\begin{figure}
  \begin{equation}
    \mathit{maxDest}(obj) = 
    \begin{cases}
      \mathit{destClass}(\mathit{obj}) & \text{if $\mathit{obj}$ is a meta class} \\
      \mathit{obj} & \text{if $\mathit{obj}$ is a non-meta class} \\
      \mathit{class}(\mathit{obj}) & \text{else}
    \end{cases}
  \nonumber
  \end{equation}

  \begin{equation}
    \mathit{maxGen}(\mathit{obj}) = \#\mbox{helix classes} + \mathit{between}(\mathit{maxDest}(\mathit{obj}), \mbox{\lstinline{Object}}) + 2
    \nonumber
  \end{equation}

  \caption{Maximum number of singleton class generations when accessing a singleton class. $\mathit{between}(A, B)$ be the number of classes between $A$ and $B$ in $A$'s superclass hierarchy. In the formula, we have to add $2$ because $\mathit{between}(A, B)$ does not count $A$ and because we have to count $obj$'s singleton class.}
  \label{fig:eq:max_singleton_classes}
\end{figure}

Consider, for example, that we want to access the third-level singleton class for the \lstinline{SmallInteger} instance \lstinline{42} and that the second-level singleton class was already generated. We may have to generate \lstinline{42 class class class}, \lstinline{SmallInteger class class}, second-level singleton classes for \lstinline{SmallInteger}'s superclasses until \lstinline{Object} (3 classes in GemStone/S), and second level singleton classes for all helix classes (5 classes in Maglev). In total, we may have to generate $10$ singleton classes or less.

We think that the number of singleton classes that must be generated, is manageable and will not have a major effect on MagLev's performance, because it is unusual to generate many high-level singleton classes and singleton class generation usually takes place directly after parsing Ruby source code files~\cite{gunter:metaprog}.

\paragraphIndent{Persisting Singleton Classes} Classes are first-class objects in GemStone/S, so we can persist them just like any other object. The persistence concept in GemStone/S is \emph{Persistence by Reachability}. When an object is persisted, the virtual machine does not only persist all instance variables but also the virtual class reference and the superclass reference if it is a class. MagLev does currently not support persisting changes that occurred in the process of singleton class generation, i.e. we can generate singleton classes for persisted objects but not persist the new reference to the singleton class object. This would require changing the persisting procedure in the virtual machine in such a way that the singleton class and its destination class are added to the dirty objects list.

\section{Ruby Modules}

Modules are the Ruby implementation of mixins, a way to add additional behavior to classes without using multiple inheritance~\cite{Bracha:1990:MI:97945.97982}. We can define modules in the same way as we can define classes and they can have singleton classes, too. We can include modules in classes, making their module methods available as instance methods of the class. If multiple included modules define the same method then the last method definition will overwrite the previous ones.

\paragraphIndent{Method Lookup} Suppose, the object \lstinline{a} is an instance of class \lstinline{A} and we first included the module \lstinline{M1} and afterwards the module \lstinline{M2} in \lstinline{A}. When we call a method, Ruby looks for the method in the following order: \lstinline{a}'s singleton class, \lstinline{A}, \lstinline{M2}, \lstinline{M1}, \lstinline{A}'s superclass. From there on, Ruby keeps looking at the next superclass and its included modules until the next superclass is \lstinline{nil}. If no method was found, the Ruby method \lstinline{method_missing} is invoked. \lstinline{super} calls within module methods or instance methods are delegated to the next object in the list.

\subsection{Problem}

We consider Ruby modules implemented correctly, if the following features are supported by MagLev and the virtual machine.

\begin{itemize}
  \item Defining modules and including modules in classes from Ruby code.
  \item Module-aware method lookup.
  \item Compatibility to exisiting Smalltalk code. The class hierarchy should not be changed because some Smalltalk applications might make assumptions about instance-of and superclass relations.
\end{itemize}

\subsection{Solution}
MagLev's Ruby parser was taken from Ruby MRI and adapted to generate MagLev AST nodes. Now, we have to elaborate how to process AST nodes involving modules -- how to represent modules in MagLev and how to include modules into classes.

\paragraphIndent{Representing Modules}
In MagLev, a module is a class that cannot have instances. A module is an object that has the class \lstinline{Module}, i.e. its virtual class' first non-meta superclass is \lstinline{Module}. It can have arbitrarily many levels of singleton classes. As we can see in Figure~\ref{fig:prob_classmodel_smalltalk_meta}, \lstinline{Module} is the superclass of \lstinline{Metaclass3}. A module's superclass is always \lstinline{Object}. \lstinline{Module} provides functionality for instance/module methods, instance variables and constants handling. In MagLev, modules are regarded as classes that cannot be instantiated.

\begin{figure}
  \centering
      \subfloat[The programmer's point of view: including modules in classes \lstinline{Object} and \lstinline{Array}.]{\includegraphics[scale=0.54]{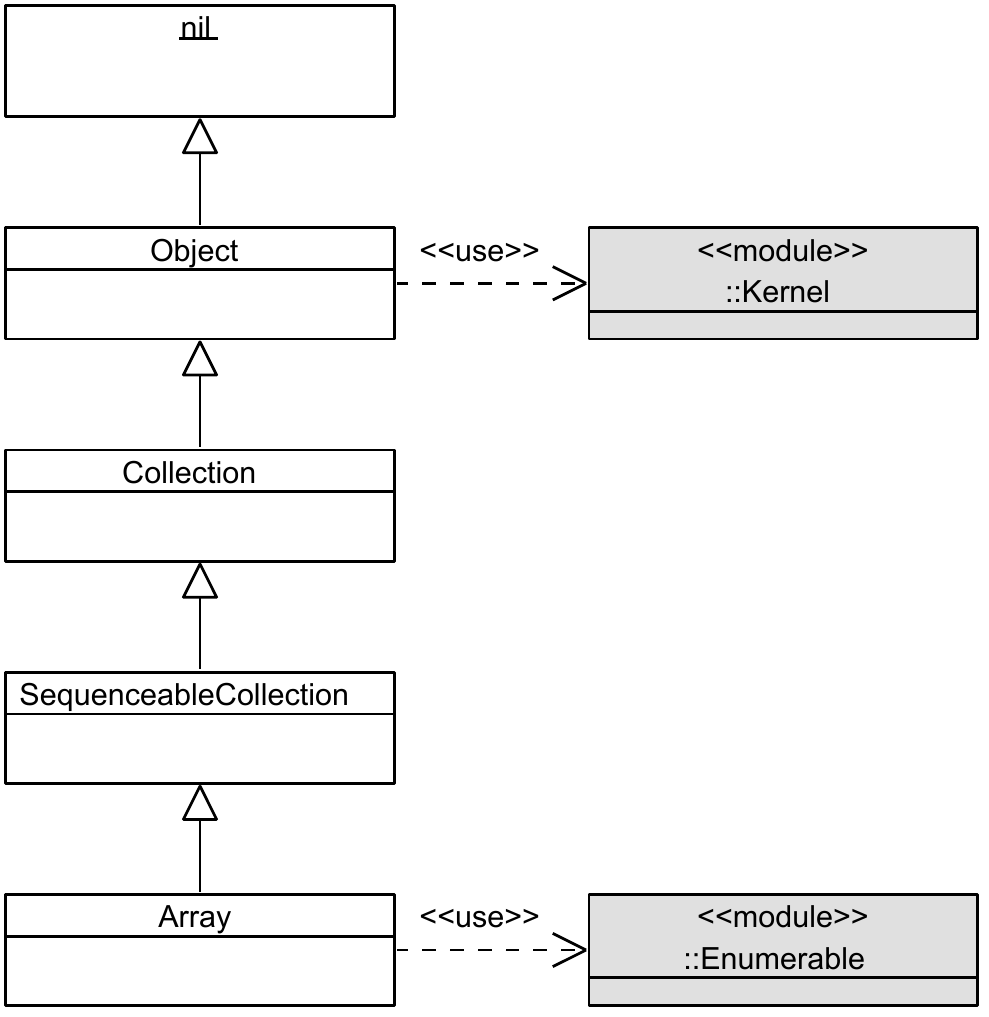}}\qquad
      \subfloat[Implementation: adding module classes to the class hierarchy.]{\includegraphics[scale=0.54]{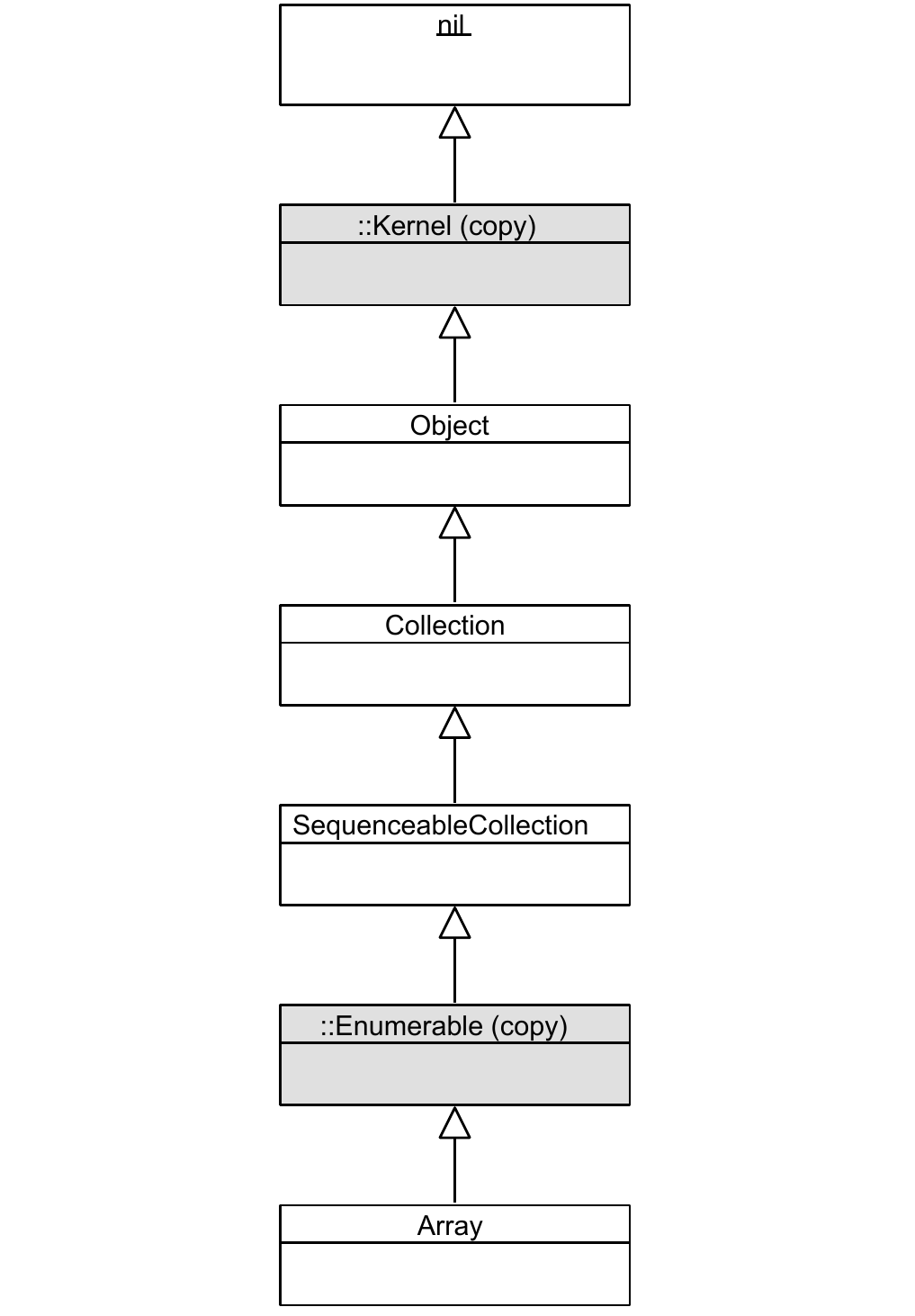}}\qquad
   \caption[Titel des Bildes]{Implementing Ruby modules by inserting copies into the superclass chain. Modules are colored gray.}
   \label{fig:ruby_maglev_modules}
\end{figure}

\paragraphIndent{Including Modules}
In order to simulate modules in an object-oriented programming language without modules or mixins, we add included modules as superclasses\footnote{Bracha and Cook call mixins ``abstract subclasses''~\cite{Bracha:1990:MI:97945.97982}. In Ruby, we regard them as abstract superclasses, because instance methods take precedence over module methods.} to the class hierarchy. Figure~\ref{fig:ruby_maglev_modules} shows Ruby modules from the programmer's point of view and their implementation as superclasses.

The process of including a module \lstinline{M} in the class \lstinline{C} involves the following steps.

\begin{enumerate}
  \item Create a class \lstinline{P} that contains all module methods as instance methods.
  \item Set \lstinline{P}'s superclass to \lstinline{C}'s superclass.
  \item Set \lstinline{C}'s superclass to \lstinline{P}.
\end{enumerate}

It is important that we operate on copies of \lstinline{M} because we can include \lstinline{M} in more than just one class. In this case, the copies of \lstinline{M} must have different superclasses.

\subsection{Implementation}
On the implementation level, MagLev's way of handling modules has some characteristics.

\paragraphIndent{Superclass References}
In MagLev, every class can have a different superclass for each environment. For example, \lstinline{Array}'s Smalltalk superclass is \lstinline{Sequencable Collection} whereas its Ruby superclass \lstinline{::Enumerable} (a module). These superclass references are stored in \lstinline{Behavior}'s instance variable \lstinline{methDicts}. This is an array that contains method dictionaries and superclass references for every environment. The superclass that is used during method lookup is the superclass in the environment of the currently executing method, i.e. the Ruby superclass is used in Ruby methods and the Smalltalk superclass is used in Smalltalk methods.


Some classes that are usually not accessed from the Ruby environment, for example GemStone/S GCI classes, have only the Smalltalk method dictionary instead of this array. In this case, the superclass reference that is stored in \lstinline{Behavior}'s instance variable \lstinline{superClass} is used during method lookup.

\begin{figure}
\begin{center}\includegraphics[scale=0.54]{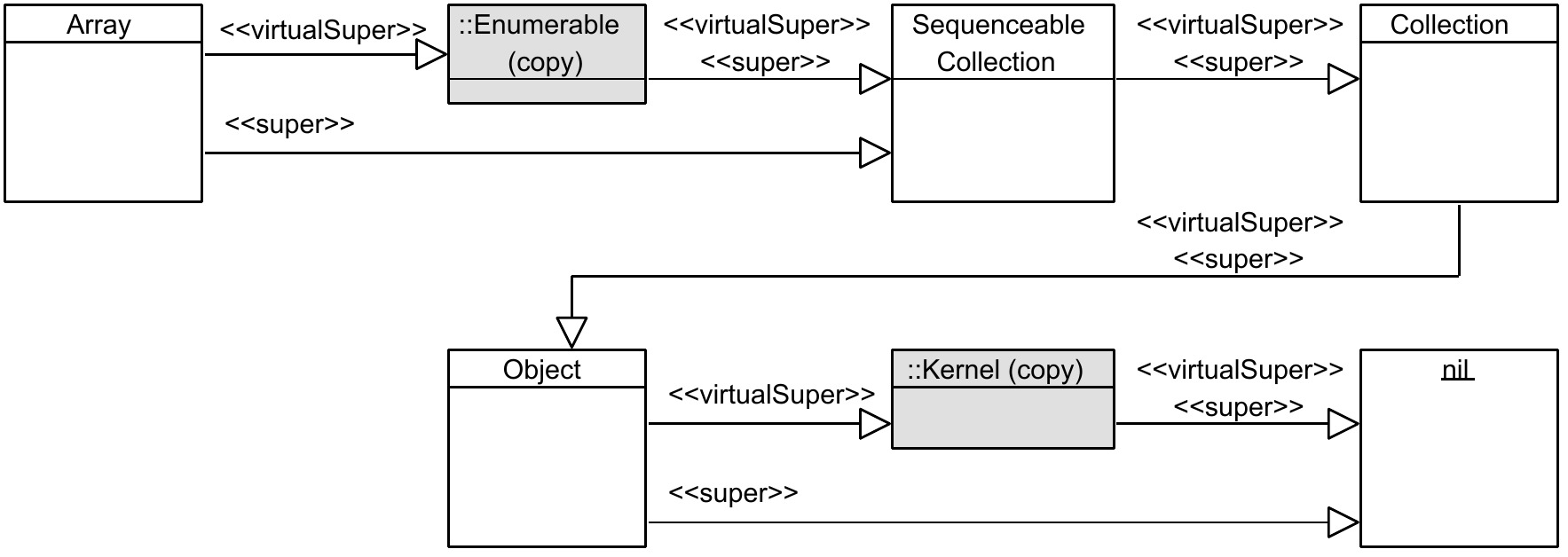}\end{center}
\caption{Superclasses and virtual superclasses for \lstinline{Array}'s class hierarchy. Virtual superclasses are colored gray.}
\label{fig:virtual_superclass}
\end{figure}

\paragraphIndent{Virtual Superclasses}
In MagLev, copies of modules that were inserted in the superclass hierarchy are called virtual classes\footnote{Not to be confused with the virtual class pointer.}. The virtual superclass is the actual superclass of the class. However, when we ask a class for its (non-virtual) superclass, we get the first virtual superclass that is not virtual. For example, in Figure~\ref{fig:virtual_superclass}, \lstinline{Array}'s virtual superclass is \lstinline{::Enumerable} but its superclass is \lstinline{SequencableCollection}. The virtual superclass is used during method lookup. The flag that determines whether a class is virtual is set during module inclusion.

All copies of modules that were included in classes are virtual classes in \mbox{MagLev}. Therefore, we can easily implement functionality that needs to distinguish between actual classes and included modules in the class hierarchy. For example, the implementation of \lstinline{included_modules} selects only virtual classes in the superclass hierarchy.

\subsection{Evaluation}
The MagLev implementation solves all the problems presented before. We can define modules in the Ruby code, include modules in classes in the Ruby code, and the method lookup is module-aware. Legacy Smalltalk code is not affected by included modules because the Smalltalk environment has its own superclass. 

However, mixins are a useful feature for the Smalltalk world, too~\cite{Cassou:2009:TWD:1412746.1412795}. It is currently not possible to define modules in the Smalltalk environment. Furthermore, there is no convenient way of adding Smalltalk methods to existing modules. Existing modules do not have a Smalltalk name because they were defined in Ruby. Therefore, they are not listed in the class browser.

More work needs to be done to provide a way for defining modules in Smalltalk. For example, MagLev could provide a method \lstinline{subclassModule:} that must not be sent to classes other than \lstinline{Object} and generates a class with the (non-virtual) class \lstinline{Module}. To include modules from Smalltalk, MagLev could provide \lstinline{subclass:} methods with an additional \lstinline{includedModules: anOrderedCollection} parameter, similar to the \lstinline{uses:} parameter in Pharo's traits implementation~\cite{Scharli03traits:composable}.

\section{Inter-language Method Invocation}
\label{sec:inter_lang_methods}
In this section, we present how we can call Smalltalk methods from Ruby and vice versa. We start by analyzing the differences between Smalltalk and Ruby regarding method calling.

\paragraphIndent{Smalltalk Methods}
In Smalltalk, we know the number of parameters for a method when we look at the selector. The number of parameters is encoded in the selector string and is equal to the number of colons. Therefore, it is not possible to call a method with the wrong number of arguments. Smalltalk does not support the concept of method visiblity: all methods are public.

\paragraphIndent{Ruby Methods}
A Ruby selector does not indicate the number of parameters. Furthermore, Ruby supports optional parameters, splat parameters and every method can implicitly take a block argument. Ruby supports three visibility modes for methods: public, protected, and private. Methods are public by default. 

\subsection{Problem}
The following features should be supported by MagLev and the \mbox{GemStone/S} virtual machine.

\begin{itemize}
  \item Adding Ruby and Smalltalk methods to an existing class.
  \item Calling Ruby methods from Smalltalk and Smalltalk methods from Ruby.
  \item Supporting method visibility for Ruby methods. For example, calling a private Ruby method from Ruby or Smalltalk should raise an exception.
  \item Raising an \lstinline{ArgumentError} when calling a Ruby method from Ruby or Smalltalk with too few or too many arguments.
\end{itemize}

\subsection{Solution}
In dynamically typed languages, methods are typically stored in a \emph{method dictionary} that maps selectors to methods. In MagLev, every environment has its own method dictionary. This is necessary because Ruby and Smalltalk define different methods with the same selector. For example, in Ruby, \lstinline{'A' * 3} returns \lstinline{'AAA'} whereas the same expression produces a \lstinline{MethodNotUnderstood} exception in GemStone/S.

When we send a message to an object, the virtual machine looks up the selector in the method dictionary of the sending environment. For example, when we call a method from Ruby, MagLev looks for the selector in the Ruby method dictionary. We need special constructs to call methods in another programming languages.

\paragraphIndent{Ruby Method Syntax} To call Ruby methods from Smalltalk, the virtual machine must perform the lookup in the Ruby method dictionary instead of the Smalltalk method dictionary. MagLev extends the Smalltalk syntax such that Ruby selectors can be called.

Ruby selectors start with \lstinline{@ruby1:} and additional parameters are added with \lstinline{_:}. Instead of the underscore we can write any other string that is allowed as part of a Smalltalk selector. Optional arguments are treated like normal arguments, as we can see in Figure~\ref{fig:call_ruby_from_st}.

\begin{figure}[htbp]
  \begin{minipage}[b]{0.475\textwidth}
    \lstset{basicstyle=\ttfamily,language=ruby} 
    \begin{lstlisting}
class Person
  def self.new(block)
    block.call(super)
  end

  def set_name(last, first = '')
    @first = first
    @last = last
  end

  def full_name
    [@first, @last].join(' ')
  end
end
    \end{lstlisting}

  \subfloat[Ruby data model for a person. The constructor creates an instance and calls the block, if given.]{\label{fig:call_ruby_from_st_ruby}\makebox[\textwidth]{}}
  \end{minipage}
  \hfill
  \begin{minipage}[b]{0.475\textwidth}
    \lstset{escapeinside={!}{!}}
    \begin{lstlisting}
!\textbf{Person class}!>>dummy
  ^ self @ruby1:new: [|person|
    person 
      @ruby1:set_name: 'Doe' 
        _: 'John';
      yourself].

!\textbf{Person}!>>testFullName
  |person name|
  person := self class dummy.
  name := person @ruby1:full_name.
  self 
    assert: name
    equals: 'John Doe'.

    \end{lstlisting}

  \subfloat[Test case that creates a person model object and tests its full name in Smalltalk.]{\label{fig:call_ruby_from_st_st}\makebox[\textwidth]{}\par}
  \end{minipage}
 \caption{Calling Ruby methods from Smalltalk, including optional parameters and block parameters.}
 \label{fig:call_ruby_from_st}
\end{figure}

\paragraphIndent{Ruby Primitives}
For calling a Smalltalk method from Ruby, we first have to create an entry in the Ruby method dictionary for the Smalltalk method. For this reason, MagLev provides the \lstinline{primitive}\footnote{In Smalltalk, we can use primitives to call native code. In this context, we are referring to Ruby primitives for calling Smalltalk methods.}  method. Figure~\ref{fig:call_st_from_ruby} shows how to call Smalltalk methods from Ruby.

\begin{figure}[htbp]
  \begin{minipage}[b]{0.435\textwidth}
    \lstset{escapeinside={@}{@}}
    \begin{lstlisting}
Object subclass: 'Person'
  instVarNames: @\color{MidnightBlue}{\#(first name)}@
  ...

@\textbf{Person class}@>>with: aBlock
  |person|
  person := self new.
  aBlock value: person.
  ^ person

@\textbf{Person}@>>name: last first: first
  first := first
  last := last.

@\textbf{Person}@>>fullName
  ^ first, ' ', last
    \end{lstlisting}

  \subfloat[Smalltalk data model for a person. \lstinline{with:} creates an instance and calls the block.]{\makebox[\textwidth]{}}
  \end{minipage}
  \hfill
  \begin{minipage}[b]{0.495\textwidth}
    \lstset{language=ruby}
    \begin{lstlisting}
class Person
  class_primitive '__new', 'with:'
  primitive 'set_name', 'name:last:'
  primitive 'full_name', 'fullName'

  def self.dummy
    __new do |person|
      person.set_name('Doe', 'John')
    end
  end

  def test_full_name
    name = class.dummy.full_name
    assert_equal(name, 'John Doe')
  end
end
    \end{lstlisting}

  \subfloat[Test case that creates a person model object and tests its full name in Ruby.]{\makebox[\textwidth]{}\par}
  \end{minipage}
 \caption{Calling Smalltalk methods from Ruby, using primitives. Smalltalk does not support optional parameters.}
 \label{fig:call_st_from_ruby}
\end{figure}

\lstinline{primitive} adds a Smalltalk method to the Ruby method dictionary and allows us to define a new selector for the method. \lstinline{class_primitive} operates on the singleton class.

\paragraphIndent{Bridge Methods}
In Ruby, a method call with a wrong number of arguments results in an \lstinline{ArgumentError} exception. MagLev implements this behavior at the method dispatch level. For every Ruby method, a number of bridge methods is generated. A bridge method's selector (\emph{full selector}) consists of the method's selector and a suffix that indicates number and type\footnote{Splat argument or block argument.} of the arguments.

\begin{figure}
\center
\includegraphics[]{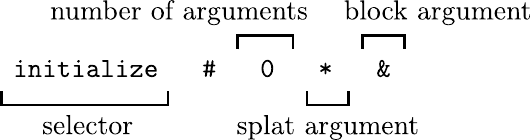}
\caption{Ruby selector with suffices for \lstinline{initialize(*args, &block)}.}
\label{fig:bridge_method_syntax}
\end{figure}

Figure~\ref{fig:bridge_method_syntax} shows the syntax for Ruby bridge method selectors. The number after the number sign is the number of arguments that the method expects. If the next character is a star, then the method takes an additional splat argument. If it does not, this character must be an underscore. The last character is an ampersand if the method takes an additional block argument. If it does not, this character must also be an underscore.

\begin{figure}
\center

\includegraphics[width=\textwidth]{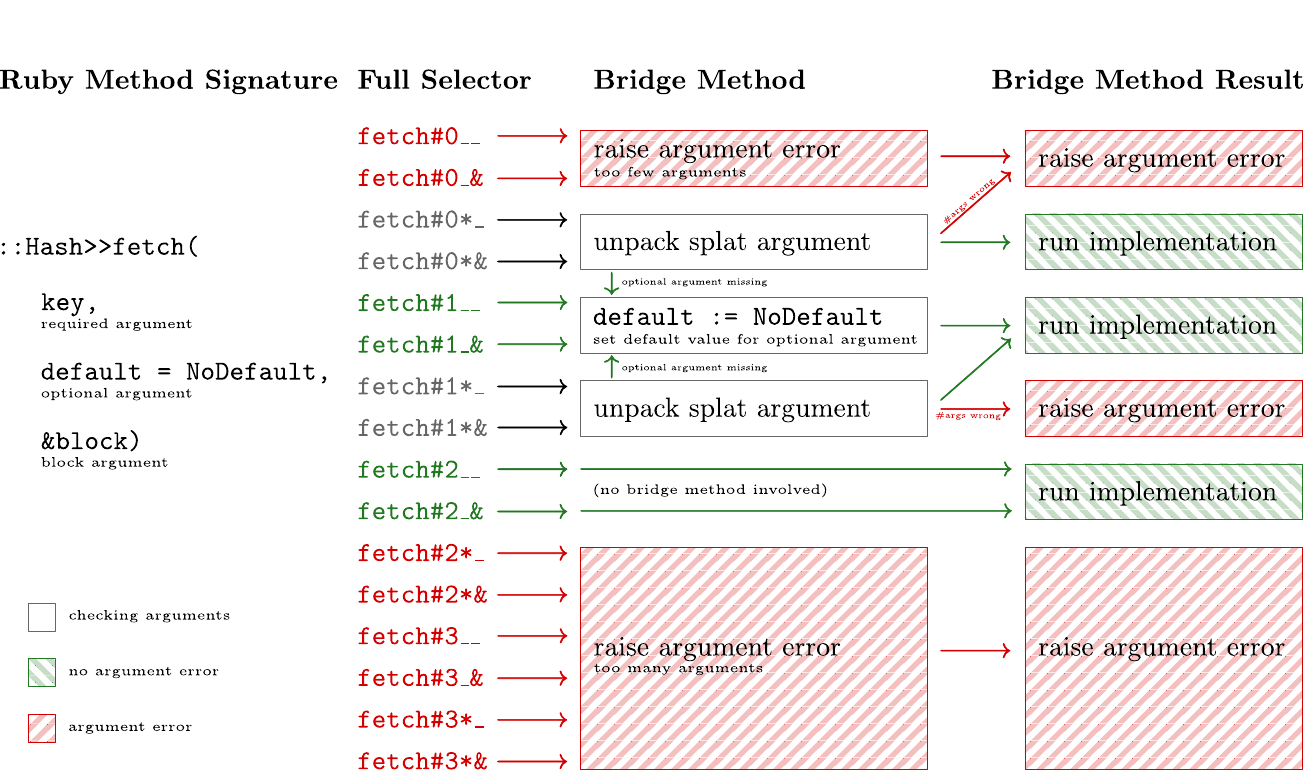}
\caption{Automatically generated bridge methods for \lstinline{::Hash>>fetch}. Methods that will raise no argument error are colored green. Methods that might raise an argument error depending on the splat argument are colored gray. Methods that will definetely raise an argument error are colored red.}
\label{fig:generated_bridge_methods}
\end{figure}

Every time we define a method, at least 16 bridge methods are generated (Figure~\ref{fig:generated_bridge_methods}): all combinations for zero to three parameters and with or without block argument or splat argument. Only the bridge method with the correct signature executes the newly-defined method. The other bridge methods raise an \lstinline{ArgumentError}. If the method takes more than three parameters, a 17th bridge method with the correct signature is generated. If a method takes optional arguments, all bridge methods that miss one or all optional arguments (e.g. \lstinline{fetch#1__}) call the \emph{full} bridge method (\lstinline{fetch#2_&}) with the maximum number of arguments and automatically provide the default argument values.

If we map a Smalltalk method to the Ruby environment with a \lstinline{primitive} call, MagLev generates bridge methods as well. MagLev does not distinguish between Ruby methods and primitives.

MagLev's Ruby compiler translates Ruby selectors in method calls to full selectors, depending on the arguments that were provided. For example, \texttt{\{\}}\lstinline{.fetch(1, 2, 3)} is translated to \texttt{\{\}}\lstinline{.fetch#3__(1, 2, 3)} and raises an \lstinline{ArgumentError}.

\subsection{Implementation}
In this section, we show some characteristics of MagLev's implementation of the concepts presented before. 

\paragraphIndent{Ruby Method Syntax}
Figure~\ref{fig:st_ruby_syntax_ebnf} shows the syntax for calling Ruby methods from Smalltalk.

\begin{figure}
\begin{verbatim}
<ruby-call-node>  ::= '@ruby1:' <selector> [<first-arg>]
<first-arg>       ::= ':' <arg-value> [<next-arg>] [<splat-arg>] [<block-arg>]
<next-arg>        ::= <normal-arg> [<next-arg>]
<normal-arg>      ::= '_:' <arg-value>
<block-arg>       ::= '__BLOCK:' <arg-value>
<splat-arg>       ::= '__STAR:' <arg-value>
\end{verbatim}
\caption{Smalltalk syntax for Ruby method calls in extended Backus-Naur form. $\langle\mbox{\emph{selector}}\rangle$ can be any valid Ruby selector. $\langle\mbox{\emph{arg-value}}\rangle$ must be a valid Smalltalk expression, additional brackets might be necessary.}
\label{fig:st_ruby_syntax_ebnf}
\end{figure}

In Figure~\ref{fig:call_ruby_from_st_ruby}, the \lstinline{block} parameter for \lstinline{self.new} was not passed as a Ruby block argument but as a normal argument. In Ruby, block parameters begin with an ampersand in the method signature.

In MagLev, it is currently not possible to call Ruby methods without normal arguments and with a block argument or a splat argument. It is not trivial to change this because we need a colon after $\langle\mbox{\emph{selector}}\rangle$. Otherwise, $\langle\mbox{\emph{block-arg}}\rangle$ or $\langle\mbox{\emph{splat-arg}}\rangle$ would be misinterpreted as a message send to the result of '@ruby1:' $\langle\mbox{\emph{selector}}\rangle$.

\paragraphIndent{Ruby Primitives}
Before we can call Smalltalk methods from Ruby, we have to add them to the Ruby method dictionary. The method \lstinline{primitive} generates bridge methods that either raise an \lstinline{ArgumentError} or contain a copy of the Smalltalk method object. \lstinline{primitive} automatically detects the number and the types of the arguments by evaluating the Smalltalk selector.

\paragraphIndent{Bridge Methods}
For every Ruby method, a number of bridge methods are generated. MagLev generates a full selector for every Ruby method call from the Ruby environment. The concept described in the solution paragraph is implemented slightly different: if we call a method with more than three arguments, MagLev always generates the full selector with three arguments and a splat argument. For instance, if we call the method \lstinline{add_numbers(1, 2, 3, 4, 5)}, MagLev will generate the full selector \lstinline{add_numbers#3*_} and pass the last two arguments as a splat argument. Therefore, the bridge method can unpack the splat argument and raise and argument error if the number of arguments does not match. If MagLev did not generate the full selector in such a way, we would get a \lstinline{NoMethodError} because MagLev generates bridge methods for only up to three arguments.

In Ruby, existing methods are overwritten if we redefine a method. Ruby does not support method overloading, i.e. we cannot define multiple methods with a different number of arguments. In MagLev, this is possible during bootstrap, when basic Ruby classes are created. For example, MagLev needs to execute different Smalltalk methods for the Ruby method \lstinline{send}. For \lstinline{send(:join)}, MagLev calls the method \lstinline{__rubySend: #join}, whereas for \lstinline{send(:join, ',')}, MagLev calls \lstinline{__rubySend: #join with: ','}. We can use this feature to implement methods that behave differently with different number of parameters, without having to check arguments in the method body explicitly. For example, \lstinline{::Hash>>fetch} has implementations for one and two arguments: with and without the default argument.

\paragraphIndent{Ruby Wrapper}
MagLev provides an easy way to work with objects that have only Ruby methods. \lstinline{RubyWrapper} is a proxy that translates all Smalltalk message sends to the proxy to Ruby message sends to the actual object. Figure~\ref{lst:ruby_wrapper_example} shows what the test case from Figure~\ref{fig:call_ruby_from_st} looks like with \lstinline{RubyWrapper}. The source code is now easier to read and Ruby methods are seamlessly integrated in the Smalltalk environment.

\begin{figure}
  \lstset{escapeinside={@}{@}}
  \begin{lstlisting}
@\textbf{Person class}@>>dummy
  ^ (RubyWrapper on: self) new: [|person|
    person set_name: 'Doe' _: 'John'].

@\textbf{Person}@>>testFullName
  |person|
  person := self class dummy.
  name := person full_name.
  self
    assert: name
    equals: 'John Doe'.
  \end{lstlisting}

  \caption{Calling Ruby methods from Smalltalk with \lstinline{RubyWrapper}.}
  \label{lst:ruby_wrapper_example}
\end{figure}

Ruby wrappers are entirely implemented in Smalltalk. A \lstinline{RubyWrapper} instance holds a reference to the actual object. \lstinline{doesNotUnderstand:} captures all Ruby message sends from the Smalltalk environment and performs a Ruby message send, using the Ruby method \lstinline{send}\footnote{\lstinline{send} is Ruby's equivalent of the Smalltalk method \lstinline{perform:}.} and the syntax shown in Figure~\ref{fig:st_ruby_syntax_ebnf}. Furthermore, \lstinline{RubyWrapper} automatically wraps all return values and block arguments before they are passed to a Smalltalk block.

\paragraphIndent{Method Visibility}
A method's visibility is saved inside an instance variable for the method object. Method visibility is checked during method lookup. The method lookup algorithm evaluates the method visibilty flag and the calling method. If the algorithm determines that the method must not be called, it returns no method object -- just as if no method was found at all.

In that case the method \lstinline{method_missing} is called. This method generates a \lstinline{NoMethodError}. It also checks if there is a private or protected method with that name and provides an exception description text accordingly.

\subsection{Evaluation}

MagLev implements the features that we asked for in the problem section: we can add and call methods in another environment, and MagLev respects method visibility. Bridge methods ensure that MagLev raises an argument error if the number of arguments is wrong. However, we noticed that existing GemStone/S tools were not made for MagLev and lack Ruby support.

\paragraphIndent{Adding Methods}
It is possible to add Smalltalk methods and Ruby methods to existing classes. In Ruby, we simply open the class again and add the method definition (also called \emph{monkey patching}). In Smalltalk, we can add new methods to existing classes in the class browser or use \lstinline{Behavior>>compile:}. GemStone/S does not come with a graphical user interface. Therefore, we have to use a frontend that connects to the stone via network (GCI), e.g. GemTools, an IDE written in Squeak.

It is, however, difficult to add Smalltalk methods to classes that do not have a Smalltalk name (\emph{nameless classes}), i.e. classes that do not have an entry in Smalltalk's globals dictionary. By default, this applies to all classes that were defined in Ruby. Existing GemStone/S IDEs such as GemTools and the VisualWorks GemStone frontend were not designed to work with MagLev: the class browser does not show nameless classes and modules. We developed the MagLev Database Explorer\footnote{\url{https://github.com/matthias-springer/maglev-database-explorer-gem}}, an experimental MagLev IDE, that solves this problem.

\paragraphIndent{Calling Methods}
We can call Ruby methods from Smalltalk and Smalltalk methods from Ruby. However, the syntax for calling Ruby methods from Smalltalk has some limitations, so that we cannot call Ruby methods with a block or splat argument without normal arguments from Smalltalk. \lstinline{RubyWrapper} solves this problem but it is inefficient. More work could be done to make \lstinline{RubyWrapper} more efficient, e.g. by providing a custom method lookup routine with a meta-object protocol~\cite{conf/tools/VranyKG12}. Furthermore, a new syntax for calling Smalltalk methods from Ruby would be convenient, such that we do not have to define primitives anymore.

\paragraphIndent{Bridge Methods}
MagLev generates bridge methods for up to three arguments when a Ruby method is defined. Therefore, when we call a Ruby method from the Smalltalk side with the wrong number of arguments, we get a \lstinline{MethodNotUnderstood} exception if we use more than three arguments. Further work needs to be done to raise an \lstinline{ArgumentError} instead. One approch is to always call the bridge method with a splat argument when calling a Ruby method with more than three parameters from Smalltalk. MagLev uses the same technique for calling Ruby methods from Ruby.

\section{Accessing Instance Variables}
\label{sec:instance_variables}
Ruby and Smalltalk have different concepts of instance variables. In Smalltalk, instance variables have to be defined as part of the class definition. Therefore, if we define new instance variables, the class object must be updated and all instances must be migrated to the new class version. In Ruby, we can dynamically add new instance variables at runtime. We do not have to specify them on class creation. We can simply access instance variables in the source code without defining them anywhere else.

\subsection{Problem}
The following features should be supported by MagLev and the \mbox{GemStone/S} virtual machine.

\begin{itemize}
  \item Defining instance variables on class definition in Smalltalk.
  \item Defining instance variables at any time in Ruby.
  \item Accessing instance variables defined in Ruby or Smalltalk from the other programming language.
  \item Having different instance variables for different objects of the same class.
\end{itemize}

\subsection{Solution}
MagLev introduced the concept of dynamic instance variables. In addition to \emph{normal} (\emph{static}) instance variables that are defined on class creation, dynamic instance variables can be defined at any time.

\paragraphIndent{Static Instance Variables}
If we define an instance variable in the Smalltalk class definition, a static instance variable is created. Therefore, we can define static instance variables in Smalltalk only. We can use static instance variables in Ruby and Smalltalk like normal instance variables. 

\paragraphIndent{Dynamic Instance Variables}
If we define or access an instance variable that was not defined in the Smalltalk class definition in Ruby, MagLev accesses a dictionary that contains the dynamic instance variables for that object. Every object has its own dictionary. Therefore, two different objects that are instances of the same class can have different instance variables. This is not possible with static instance variables because they must be defined in the class definition. 

If we want to access dynamic instance variables in Smalltalk, we have to use \lstinline{dynamicInstVarAt:} and \lstinline{dynamicInstVarAt:put:}. In Smalltalk, we have to know whether an instance variable is static or dynamic because we have to use different constructs to access them.

\subsection{Implementation}
In Ruby, static instance variables have two names: the name of the static instance variable as it was defined in the class definition and that same name with an \lstinline{_st_} prefix. For instance, we can access the static instance variable \lstinline{size} in \lstinline{::Hash} with \lstinline{@size} and \lstinline{@_st_size}. If we call \lstinline{instance_variables} in Ruby, we only get the prefixed names. This is the only way to distinguish static instance variables from dynamic instance variables in Ruby. MagLev provides a way to hide specific static instance variables from Ruby. For example, \lstinline{destClass} in \lstinline{Metaclass3} is not visible in the Ruby environment.

\subsection{Evaluation}
MagLev's implementation of dynamic and static instance variables solves all requirements that we listed in the problem section. Smalltalk and Ruby programmers can define instance variables in the way they are used to. With dynamic instance variables it is possible to add new instance variables at any time, in both Ruby and Smalltalk. Futhermore, it is possible to use instance variables beyond environment boundaries. However, dynamic instance variables should be integrated more seamlessly in Smalltalk.

\paragraphIndent{Integration of Dynamic Instance Variables}
In Ruby, dynamic and static instance variables are integrated seamlessly in the programming language. As a programmer, we do not have know if we are accessing a dynamic or static instance variable. In Smalltalk, however, we need to use special constructs to access dynamic instance variables. In future versions of MagLev, the Smalltalk compiler could be adapted, such that all references to dynamic instance variables are implicitly replaced by the method calls for reading and writing dynamic instance variables.

\section{Related Work}
\label{sec:related-work}
In section, we compare the solutions presented before and their implementation in MagLev to other multi-language virtual machines. We will see that some virtual machines and languages were specifically designed to support multiple languages, whereas others were not and had to solve problems that are similar to ours.

\subsection{.NET CLI Languages}
The Common Language Infrastructure~(CLI), standardized as \mbox{ECMA-335}~\cite{Ecma335}, is an open specification for language-independent and platform-in\-de\-pen\-dent software development. Among other things, it describes the Virtual Execution System~(VES), the Common Language Specification~(CLS), and the instruction set of the Common Intermediate Language~(CIL). The CLI was designed to support multiple programming languages. Compilers transform source code to CIL code and the VES executes that code. 

\paragraphIndent{Common Language Specification}
The CLS is a set of rules that is important for language implementors and application developers. Some programming languages offer features that are not supported in other programming languages. For example, according to Hamilton, ``incompatible types are the primary barriers that keep languages from interoperating''~\cite{Hamilton:2003:LIC:772970.772973}. The CLS ensures that CLS-compliant components can interact with each other, e.g. by defining a set of data types that all CLS-compliant languages support. Language implementors have to support these data types to call their implementation CLS-compliant and their compilers must generate CLS-compliant artifacts that do not use other data types in their public interface. 

\paragraphIndent{CLI classes}
In contrast to Smalltalk, it is not possible to add new methods to existing CLI classes. Therefore, we can write source code for a class in only one language. For example, it is not possible add methods written in both C\# and VB.NET to the same class.

\subsection{C\# and VB.NET}
The two most popular languages for the .NET framework are C\# and Visual Basic .NET. Both languages were developed for the CLI. They both use the same object model and the same standard library. The .NET framework does not support mixing different languages in one assembly\footnote{Assemblies are CIL artifacts, i.e. EXE files or DLL files.}. The only way to mix both languages is to create an assembly/library in one language and to import it in an assembly written in the other language. 

\paragraphIndent{CLS Compliance}
The public interface of an assembly should be CLS-compliant. Consider, for example, that we created a C\# assembly with a class that has the instance methods \lstinline{example} and \lstinline{eXample}. From VB.NET, we cannot call any of these methods because VB.NET is case-insensitive. According to the CLS, ``for two identifiers to be considered different under the CLS they shall differ in more than simply their case.''~\cite{Ecma335}.

\paragraphIndent{Conclusion}
For C\# and VB.NET, the object model does not have to be mapped or transformed because both languages have the same object model. Both languages share the same type system. Components written in VB.NET and C\# can interact with each other as long as they are CLS-compliant.

\subsection{Ruby .NET}
The first implementation of Ruby for the CLI was Ruby .NET\footnote{\url{https://code.google.com/p/rubydotnetcompiler/source/checkout}}, developed at the Queensland University of Technology. 

\paragraphIndent{Architecture}
In Ruby .NET, Ruby objects are always instances of \lstinline{Ruby.Object}. Every Ruby object has a reference to its class and a dictionary that contains the instance variables. Futhermore, Ruby .NET has its own \lstinline{Ruby.Class} to support instance method changes at runtime and other Ruby features.

\paragraphIndent{Calling Methods}
In Ruby .NET~\cite{Kelly:2008:RRC:1378279.1378288}, instances of \lstinline{Ruby.Class} have their own method dictionaries that map selectors to Ruby methods. Ruby methods are represented as singleton\footnote{We refer to the singleton design pattern~\cite{GlossarWiki:Gamma_et_al.:1995}. Not to be confused with Ruby singleton classes.} classes with multiple \lstinline{call} methods for up to 10 parameters and a method for more than 10 parameters: \lstinline{call0}, \lstinline{call1}, \lstinline{call2}, \ldots. Every method takes a block argument that may be \lstinline{null} if no block was given. The method \lstinline{call} is used for methods with a splat argument. In this case, no other arguments except for the block argument are allowed. The abstract Ruby method superclass raises an argument error for all \lstinline{call} methods. Ruby method classes overwrite \lstinline{call} methods that have the correct number of arguments and execute the actual method. 

When an instance of a CLI class is accessed in Ruby for the first time, Ruby .NET ``dynamically create[s] a special \lstinline{Ruby.Class} object to represent the foreign CLI type.''~\cite{Kelly:2008:RRC:1378279.1378288}. These Ruby classes use CLI reflection instead the Ruby method dictionary during method lookup. Ruby .NET retains a dictionary that maps CLI classes to already generated Ruby classes.

If we want to call methods on a CLI object, we do not need a special syntax. We can call methods on CLI objects like any other Ruby method. Calling a Ruby method from another CLI language is more difficult. We have to use the method \lstinline{Ruby.Eval.CallPublic} and provide the Ruby object, a caller context and the arguments of the Ruby method. Ruby .NET provides property getters for some important Ruby objects, e.g. \lstinline{Ruby.Inits.rb_cObject} returns the Ruby \lstinline{Object} class.

\paragraphIndent{Ruby Modules}
The CLR does not support mixins. Modules and classes are represented as the same CLI classes \lstinline{Ruby.Class}. This class has a type instance variable that determines if the object is a class or a module. Depending on that flag, certain methods raise an error or behave differently. When a module is included in a class, Ruby .NET creates a new class, copies over all instance methods and instance variables to the class, sets the type variable accordingly, and changes the superclass references, similarly to MagLev.

\paragraphIndent{Ruby Singleton Classes}
Ruby .NET generates singleton classes on the first access. It does not support generating singleton classes higher than first-level singleton classes. Higher-level singleton classes can be generated, but their superclass is not set correctly.

\paragraphIndent{Comparison to MagLev}
The biggest conceptual difference between MagLev and Ruby .NET is that, in MagLev, Ruby classes and Smalltalk classes are the same objects. In Ruby .NET, there is a CLI class and a Ruby class for some classes. Classes that were defined in Ruby do not have a CLI class at all. Ruby method selectors are implemented similarly in Ruby .NET and MagLev. In both implementations, methods exist for different argument numbers and argument types. In MagLev, we call these methods bridge methods. Modules and singleton classes are also implemented similarly. In Ruby .NET, however, they are entirely built on top of the CLR: Ruby .NET simulates the method lookup for Ruby objects, whereas, in MagLev, the virtual machine was changed to support the Ruby method lookup. 

\subsection{IronRuby}
IronRuby\footnote{\url{http://www.ironruby.net/}} is another implementation of the Ruby programming language, built on top of the Dynamic Language Runtime (DLR)~\cite{dlrspec}. The DLR is a library built on top of the CLI. It provides functionality for dynamic method dispatch, code generation, and other features of dynamically typed programming languages. All DLR functionality can be manually implemented. However, the DLR simplifies language implementation and interaction between two DLR languages (e.g. IronRuby and IronPython). The DLR also simplifies the interaction of C\# with DLR languages in combination with C\#'s \lstinline{dynamic} keyword~\cite{Bierman:2010:ADT:1883978.1883986}. This keyword was introduced in C\# 4.0 to bypass static type checking. We do not discuss IronRuby or the DLR any further in this work and encourage the reader to take a look at the DLR documentation~\cite{dlrspec}.

\subsection{JRuby}
The idea behind the Java Virtual Machine (JVM) is to support platform-in\-de\-pen\-dent software development (``write once, run anywhere'') by providing JVM implementations for many platforms. The JVM was built to support a single language: Java. Other programming languages have been implemented on top of the JVM, e.g. Scala, Groovy, and Ruby. JRuby is an implementation of Ruby for the JVM.

In JRuby, all Ruby objects are instances of the Java class \lstinline{RubyObject} and Ruby classes are instances of the Java class \lstinline{RubyClass}. Ruby classes can have only Ruby methods and Java classes can have only Java methods.

\paragraphIndent{Calling Ruby and Java Methods}
If we want to call a Ruby method from Java, we need a helper object that provides functionality for operating with Ruby source code and Ruby objects, e.g. an instance of \lstinline{ScriptingContainer}. Other helper classes exist that use the \emph{Scripting for the Java Platform API} (JSR~223\footnote{\url{http://www.jcp.org/en/jsr/detail?id=223}}). The helper object provides methods for evaluating Ruby source code, as well as reading and writing local and global variables. In addition, we can evaluate Ruby code on a Ruby object that we received at the end of a prior Ruby code evaluation. 

If we want to use a Java class from Ruby, we have to require the package \lstinline{java}. Afterwards, we can access all Java classes in the class path like normal Ruby classes~\cite{nutter2011using}. JRuby automatically wraps Java classes in a proxy object. This allows Ruby programmers to define additional Ruby methods on the wrapper object. However, these methods are not available from Java.

Every time we pass an argument to a method in the other programming language, it is converted to the according Ruby or Smalltalk class. For example, \lstinline{java.lang.Integer} objects are converted to \lstinline{Fixnum} objects and vice versa.

\paragraphIndent{Ruby Modules}
Similarly to MagLev, JRuby adds included modules as superclasses to the class hierarchy. JRuby creates so-called \emph{included module wrappers}. An included module wrapper is a \lstinline{RubyClass} that holds a reference to the module and delegates constant handling, method handling and instance variable handling to the module.

\paragraphIndent{Instance Variable Tables}
JRuby stores instance variables in variable tables. Every Ruby object has its own variable table. It consists of an array of instance variable values (Ruby objects). In addition, every Ruby object has a variable table manager that maintains the mapping of instance variable names to variable table offsets. This is similar to dynamic instance variables in GemStone/S, except that dynamic instance variables map instance variable names to values directly.

\subsection{STX:LIBJAVA}
With STX:LIBJAVA\footnote{\url{https://swing.fit.cvut.cz/projects/stx-libjava}}, Kurs et al. implemented the Java programming language on top of the Smalltalk/X virtual machine. STX:LIBJAVA~\cite{Hlopko:2012:ISJ:2448963.2448968} uses a modified Smalltalk/X virtual machine that can execute both Smalltalk byte code and Java byte code. It distinguishes between Java objects and Smalltalk object, but both live in the same object memory. When the control flow crosses the language boundary, i.e. a Smalltalk method is called from Java or the other way around, STX:LIBJAVA generates a proxy method that performs the method resolution, transforms arguments, and passes the control flow to the real method. In \mbox{MagLev}, bridge methods are a similar concept but they do not lookup methods by themselves or transform objects between programming languages. 

In a later paper, Kurs et al. evaluate the concept of behavior objects for Java in Smalltalk/X~\cite{DBLP:conf/dateso/KursVB11}. Just as in MagLev, every Smalltalk/X object is a Java object. Instead of environments, Kurs et al. use different behavior objects for every programming language. A behavior object contains the methods for an object in one programming language. A Smalltalk behavior object is a Smalltalk class and a Java behavior object is an object similar to a Java class. Therefore, objects can have two classes. In addition, they use mapping functions to transform the object layout for object state, assuming that different programming languages expect different object state layouts. Just as in MagLev, every programming language can define a different superclass. The superclass is stored in the behavior object.

\section{Conclusion}
\label{sec:conclusion}
In the work, we showed how MagLev was implemented on top of the GemStone/S virtual machine and how Ruby source code can interact with Smalltalk source code. We showed how MagLev maps the Ruby and the Smalltalk object model: MagLev extends Smalltalk's meta class model to Ruby's singleton class model and supports Ruby modules by inserting virtual classes into the superclass hierarchy. We showed how methods are invoked: MagLev introduces the concept of language environments that seperate the Ruby world from the Smalltalk world. Ruby method calling conventions and error handling is implemented with bridge methods. Finally, we showed how instance variables are stored with dynamic instance variables: they are stored in an instance variable dictionary for every object. 

MagLev's key characteristic is that every Ruby object is a Smalltalk object and vice versa. We think that this is the best form of language integration. When we looked at JRuby and Ruby .NET, we realized that an object is always either a Ruby object or a Java/CLI object. Therefore, objects must be converted or wrapped when calling into another language. MagLev can avoid this overhead.

When we analyzed related work, we realized that there are virtual machines and programming languages that were specifically designed to support and interact with multiple programming languages. Examples are the CLR and the programming languages C\# and VB.NET. These programming languages avoid some of the problems we encountered with MagLev by setting up a standard for inter-language collaboration: the Common Language Specification (CLS). However, the CLS also restricts programmers: they may not be allowed to use all language features when writing components that interact with components written in another language. We think that this is reasonable in most cases. For example, most programmers will probably never want to call methods on higher-level singleton classes in MagLev's Smalltalk environment, because singleton classes are a Ruby characteristic. Therefore, we think that we do not only ``need more open, language independent virtual machines''~\cite{vranyj2010}, but also standards like the Common Language Specification (CLS) for better language integration.


\bibliographystyle{ACM-Reference-Format-Journals}
\bibliography{references}

\appendix

\medskip

\section{Zusammenfassung}
Unter mehrsprachigen virtuellen Maschinen versteht man virtuelle Maschinen (VMs), welche die Ausf{\"u}hrung von Quelltext in verschiedenen Programmiersprachen unterst{\"u}tzen. Eine derartige Funktionalit{\"a}t bietet eine Reihe von Vorteilen, sowohl bei der Anwendungsentwicklung, als auch bei der Implementierung von Programmiersprachen: Anwendungsentwickler k{\"o}nnen Bibliotheken verwenden, die in verschiedenen Programmiersprachen geschrieben sind. Programmierer, die eine Programmiersprache implementieren, m{\"u}ssen sich keine Gedanken mehr {\"u}ber grundlegende VM-Funktionalit{\"a}ten, wie zum Beispiel Garbage Collection oder Speicherverwaltung machen.

In dieser Arbeit stellen wir MagLev, eine Implementierung der Ruby Programmiersprache auf Basis der virtuellen Maschine von GemStone/S, vor. GemStone/S ist eine Smalltalk-Implementierung. Am Beispiel von Ruby und Smalltalk zeigen wir, welche Probleme es bei der Kommunikation zwischen Softwarekomponenten gibt, die in verschiedenen Programmiersprachen geschrieben sind. F{\"u}r diese Probleme stellen wir dann die L{\"o}sung und deren Umsetzung in MagLev vor. Wir zeigen, wie man die Ruby und Smalltalk Objektmodelle aufeinander abbildet, wobei der Schwerpunkt auf Ruby Singleton-Klassen, Smalltalk Meta-Klassen und Ruby Modulen liegt. Wir zeigen au{\ss}erdem, wie Smalltalk Quelltext von Ruby (bzw. Ruby Quelltext von Smalltalk) aufgerufen werden kann und wie der Zugriff auf Instanzvariablen funktioniert.

Wir stellen fest, dass MagLev alle diese Probleme l{\"o}st. Jedoch sind die existierenden Programmierwerkzeuge nicht an MagLev angepasst, sodass einige Probleme nur auf technischer Ebene gel{\"o}st wurden. Beispielsweise ist es mit den existierenden GemStone/S IDEs nicht m{\"o}glich, Klassen zu ver"andern, die in Ruby erstellt wurden.

Beim Vergleich von MagLev mit anderen Programmiersprachen stellen wir fest, dass zahlreiche Implementierungen wie JRuby und Ruby .NET zwischen Ruby und Smalltalk Objekten unterscheiden. Das ist bei MagLev nicht der Fall: jedes Ruby Objekt ist auch ein Smalltalk Objekt. Deshalb m{\"u}ssen bei JRuby und Ruby .NET Objekte konvertiert werden, wenn Methoden in einer anderen Programmiersprache aufgerufen werden. Bei MagLev entf{\"a}llt dieser Schritt.


\end{document}